\documentclass[bezier]{amsart}
\usepackage{placeins}
\usepackage{graphicx}
\usepackage{amsmath}
\usepackage{amssymb}
\usepackage{times}
\usepackage{color}
\usepackage{epsfig}

\begin{document}
\title{Non-equilibrium allele frequency spectra via spectral methods}
\author{Sergio Luki\'c $^1$}
\address{Sergio Luki\'c,
	Department of Genetics and BioMaPS Institute, 
	Rutgers University, 
	Piscataway NJ 08854, 
	USA}
\email{lukic@biology.rutgers.edu}
\thanks{$^1$ Corresponding author.}
\author{Jody Hey}
\address{Jody Hey,
	Department of Genetics, 
	Rutgers University, 
	Piscataway NJ 08854, 
	USA}
\author{Kevin Chen}
\address{Kevin Chen,
	Department of Genetics and BioMaPS Institute, 
	Rutgers University, 
	Piscataway NJ 08854, 
	USA}

\date{ \today}

\begin{abstract}

A major challenge in the analysis of population genomics data consists of isolating signatures of natural selection from background noise caused by 
random drift and gene flow. Analyses of massive amounts of data from many related populations require high-performance algorithms to determine the 
likelihood of different demographic scenarios that could have shaped the observed neutral single nucleotide polymorphism (SNP) allele frequency spectrum. 
In many areas of applied mathematics, Fourier Transforms and Spectral Methods are firmly established tools to analyze spectra of signals and model their 
dynamics as solutions of certain Partial Differential Equations (PDEs). When spectral methods are applicable, they have excellent error properties and 
are the fastest possible in high dimension; see \cite{numericalrecipes}.
In this paper we present an explicit numerical solution, using spectral methods, to the forward Kolmogorov equations for a Wright-Fisher process with migration of $K$ populations, influx of mutations, and multiple population splitting events.
\end{abstract}

\maketitle
\numberwithin{equation}{section}

\section{Introduction}

Natural selection is the force that drives the fixation of advantageous phenotypic traits, and represses the increase in 
frequency of deleterious ones. The growing amount 
of genome-wide sequence and polymorphism data motivates the development of new tools for the study of natural selection. 
Distinguishing between genuine selection and the effect of demographic history, 
such as gene-flow and population bottlenecks, on genetic variation presents a major technical challenge. 
A traditional population genetics approach to the problem focuses on computing
neutral allele frequency spectra to infer signatures of natural selection as deviations from neutrality.
Diffusion theory provides a set of classical techniques to predict such frequency spectra 
\cite{gutenkunst, williamson, evans}, while the connection between diffusion and the theory of Partial 
Differential Equations (PDEs) allows for borrowing well established high-perfomance algorithms from applied mathematics.

\begin{figure}[hpbt]
\begin{center}
\vspace{0.5cm}
    \includegraphics[width=5.0cm]{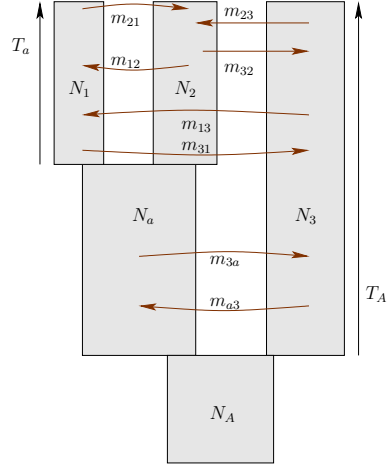}
\caption{\label{figone}
A graphical representation of a model for the demographic history of three populations.
}
\end{center}
\end{figure}

The theory of predicting the frequency spectrum under irreversible mutation was developed by
Fisher, Wright and Kimura \cite{fisher, wright, kimuraIIm}. 
In particular Kimura \cite{kimuraIII} noted that this theory was applicable to 
many nucleotide positions and introduced the \emph{infinite sites model}.
The joint frequency spectra of neutral alleles can be obtained
from the coalescent model \cite{wakeley} or by Monte-Carlo simulations \cite{hudson}. 
The analysis in terms of diffusion theory is mathematically simpler and can incorporate natural selection
easily \cite{gutenkunst, williamson, evans}.
In this paper, we model the demographic history of $K$ different populations that are descended by $K-1$
population splitting events from a common ancestral population. The populations evolve with gene exchange
under an infinite sites mutation model. We
introduce a powerful numerical scheme to solve the associated forward diffusion equations. 
After introducing a regularized discretization of the problem, we
show how spectral methods are applied to compute theoretical Non-Equilibrium Frequency Spectra.

The introduction of spectral methods is 
usually attributed to \cite{orszag}, although they are based on older precursors,
such as finite element methods, and Ritz methods in quantum mechanics \cite{ritz}. The basic idea consists of
using finite truncations of expansions by complete bases of functions to approximate the solutions of a PDE. 
This truncation allows the transformation of a diffusion PDE into a finite system of Ordinary Differential Equations (ODEs).
The motivation to use these methods relies on their excellent error properties, and their high speed.
In general, they are the preferred methods when the dimension of the domain is high \cite{numericalrecipes}, and
the solutions to the PDE are smooth. This is because the number of basis functions that one needs to approximate 
the solutions of a PDE is much lower than the number of grid-points that one needs in a finite difference scheme, 
working at the same level of accuracy \cite{spectral}.

As we show in this paper, the convergence of spectral methods depends on the smoothness of the solutions 
to be approximated. 
In many situations, solutions to diffusion equations have good analytical properties, and spectral methods
can be applied. However, the application of these methods to the problem that
interests us here requires a proper discretization of the problem. Influx of mutations, 
population splitting events and boundary conditions have to be properly regularized before one 
applies these methods and exploits their high-perfomance properties. 

\subsection{Non-Equilibrium Frequency Spectra}

The $K$-dimensional Allele Frequency Spectrum (AFS) summarizes the joint allele frequencies in $K$ populations.
We distinguish between the AFS, which consists of the unknown distribution of allele frequencies in $K$
natural populations, and \emph{observations} of the AFS. 
Given DNA sequence data from multiple individuals in $K$ populations, 
the resulting \emph{observation} of the AFS is a $K$-dimensional matrix with the allele counts (for a complete discussion
on this see \cite{wakeley}). 
Each entry of the matrix consists of the number of diallelic polymorphisms in which the 
derived allele was found. In other words, each entry of the AFS matrix
is the observed number of derived alleles, $j_a$, found in the 
corresponding number of samples, $n_a$, from population $a$ ($1\leq a\leq K$). 

The full AFS is the real distribution of joint allele frequencies at the time when the samples were collected.
If the total number of diploid individuals in the $a^{th}$-population is $N_a\gg n_a$, the natural
allele frequencies $x_1=i_1/(2N_1)$, $x_2=i_2/(2N_2)$, $\ldots x_k=i_K/(2N_K)$ 
(with $i_a$ the total number of derived polymorphisms in the $a^{th}$ population) 
can be seen as points in the $K$-dimensional cube $[0,\, 1]^K$. Thus, given 
the frequencies of every diallelic polymorphism (which we indexed by $r$) $x^r_1$, $x^r_2$,
$\ldots, x^r_K$, the AFS can be expressed as the probability density function
\begin{equation}
\phi(x)=\frac{1}{S}\sum_{r=1}^{S} \delta(x_1-x^r_1)\delta(x_2-x^r_2)\cdots\delta(x_K-x^r_K). 
\end{equation}
Here, $S$ is the total number of diallelic polymorphisms segregating in the $K$ populations, 
and $\delta(\, )$ is the Dirac delta function.

Our goal is to determine this AFS under the infinite sites model. 
Any demographic scenario in the model is defined through
a population tree topology $T$, such as in Fig. \ref{figone}, and a set of parameters that specify the effective 
population sizes $N_{e,a}$, splitting times $t_A$, and migration rates 
$m_{ab}$ at different times. Hence,
$2N_{e,a} m_{ab}$ is defined as the number of haploid genomes that the population $a$ receives from $b$ 
per generation. For simplicity, we refer to the set of parameters that specify a unique demographic
scenario as $\Theta$. 

Thus, given a population tree topology and a choice of parameters, we will compute 
theoretical densities of derived joint allele frequencies as functions on $[0,1]^K$ of the type
\begin{equation}
\label{model}
\phi(x\vert \Theta, T)=\sum_{i_1=0}^{\Lambda-1} \sum_{i_2=0}^{\Lambda-1}\cdots \sum_{i_K=0}^{\Lambda-1}
\alpha_{i_1,i_2,\ldots,i_K}(\Theta, T)R_{i_1}(x_1)R_{i_2}(x_2)\cdots R_{i_K}(x_K),
\end{equation}
with $\Lambda$ a truncation parameter, 
$\{ R_i(x) \}_{i=0}^\infty$ a complete basis of functions on the Hilbert space $L^2[0,1]$ to be defined below,
and $\alpha_{i_1\cdots i_K}$ the coefficients associated with the projection of $\phi(x\vert \Theta, T)$
onto the basis spanned by $\{ R_{i_1}(x)R_{i_2}(x)\cdots R_{i_K}(x) \}$.
These continuous densities relate to the expectation of an \emph{observation} 
of the AFS via standard binomial sampling formulae
\begin{equation}
\label{relation_AFS}
p(j_1, \ldots j_K \vert n_1,\ldots n_K)=\int_{[0,1]^{K}}  \phi(x\vert\Theta,T)\prod_{a=1}^{K}\frac{n_a!}{(n_a-j_a)!j_a!}
x_a^{j_a}(1-x_a)^{n_a-j_a}dx_a.
\end{equation}
Using Eq. \eqref{relation_AFS} we can compare model and data, for instance, by means of maximum likelihood.

The major goals of this paper are twofold. First, we present the finite Markov chain and diffusion approximation,
associated with the infinite sites model used to compute neutral allele spectra. A special emphasis is made 
on the boundary conditions and the influx of mutations, because of their potential singular behavior. Second, we 
introduce spectral methods and show
how to transform the diffusion equations into coupled systems of Ordinary Differential Equations (ODEs) that can be 
integrated numerically. In particular, we introduce a set of techniques to handle population splitting events,
mutations and boundary interactions, that protect the numerical setup against Gibbs phenomena\footnote{Gibbs phenomena
are numerical instabilities that arise when the error between a function and its truncated polynomial approximation is large.}.
A detailed analysis of the stability of the methods as a function of the model parameters, 
and the control of the numerical error, are included at the end of the paper.

\section{Finite Markov chain model}

The evolutionary dynamics of diallelic SNP frequencies in a randomly-mating diploid population 
can be modeled using a finite Markov chain, with discrete time $t$ representing non-overlapping generations.
For simplicity, we consider first one population with $N$ diploid individuals, and later will extend the
results to more than one population.

The state of the Markov chain at time $t$ is described by the vector $f_j(t)$, with $1\leq j \leq 2N$.
Each entry, $f_j(t)$, consists of the expected number of loci at which the derived state is found
on $j$ chromosomes. Therefore, $\sum_{j=1}^{2N-1}f_j(t)$ is the expected number of polymorphic loci 
segregating in the population at time $t$, and
$f_{2N}(t)$ is the expected number of loci fixed for the derived state. The model assumes
that the total number of sites per individual is so large, and the mutation rate per site so low, that
whenever a mutation appears, it always does so on a previously homoallelic site \cite{kimuraIII}.

\begin{figure}[hpbt]
\begin{center}
\vspace{0.5cm}
    \includegraphics[width=2.2cm]{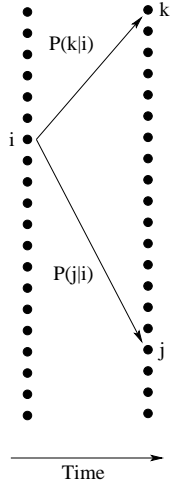}
\caption{\label{fige}
One unit of time transition in a finite Markov chain.
}
\end{center}
\end{figure}

The vector $f_j(t)$, is also called the density of states. Under the assumption of free recombination
between loci and constant mutation rate, the time evolution of $f_j(t)$ under random drift
and mutation influx is described by the difference equations
\begin{equation}\label{chain_general}
f_j(t+1)=\sum_{i=1}^{2N}P(j\vert i)f_i(t) + \mu_j(t),\quad 1\leq j \leq 2N.
\end{equation}
In its simplest form, one assumes that the alleles in generation $t+1$ are derived by sampling with 
replacement from the alleles in generation $t$. Therefore, the transition coefficients in the chain 
Eq. \eqref{chain_general} are
\begin{equation}\label{WF}
P(j\vert i)=\binom{2N}{j}(i/(2N))^j \{ 1-(i/(2N)) \}^{2N-j}.
\end{equation}
This describes stochastic changes in the state after a discrete generation, Fig. \ref{fige}.
The second term in Eq. \eqref{chain_general} represents the influx of polymorphisms. Mutations 
are responsible for the creation of new polymorphisms in the population. The influx
of mutations depends on the expected number of sites $2N\nu$, in which new mutations 
appear in the population each generation\footnote{The expected number of sites $2N\nu$, relates
to the expected number of mutations per base $2N u$, by the total length $L$ of the genomic
sequence under study in units of base pairs, $\nu=u\times L$. Sometimes in this paper, in an abuse of notation
we do not distinguish between 
$\nu$ and $u$, and they are seen as the same quantity expressed with different units.}. If we assume that at 
each generation, every new mutation is found in just one chromosome, then
\begin{equation}\label{mutation1}
\mu_j(t)=2N\nu \delta_{1,j},
\end{equation}
for the mutation alone \cite{evans}. The term $\delta_{i,j}$ in Eq. \eqref{mutation1} is the Kronecker
symbol, with $\delta_{1,j}=1$ if $j=1$ and $\delta_{1,j}=0$ otherwise.

\subsection{Effective Mutation Densities} In applications of the infinite sites model, one usually finds
that the census population size and the effective population size that drives random drift in Eq. \eqref{WF}
are not the same \cite{kimuraIII}. For this reason, we distinguish between $N_e$, the effective population size
that defines the variance of the Wright-Fisher process in Eq. \eqref{WF}, and the census population size $N$ that
can be used to define the allele frequencies $x=i/(2N)$.
Therefore, the smallest frequency, $x=1/(2N)$, with which new mutations enter populations will be 
sensitive to small stochastic fluctuations in the census population size, even if the effective 
population size remains constant. This is important when we take the diffusion limit 
of Eq. \eqref{chain_general}, and the stochastic process is described by the continuous variable
$x=j/(2N)$, rather than the integer $j$. If we consider a constant census population size, 
the term Eq. \eqref{mutation1} in the Markov chain is substituted by
\begin{equation}\label{mutation2}
 \delta_{1,j} \mapsto \delta(x-1/(2N)),
\end{equation}
in the diffusion limit. However, if the census population size per generation is a stochastic variable
distributed as $F(N)dN$, the diffusion limit of the mutation term will be
\begin{equation}\label{mutation3}
 \delta_{1,j} \mapsto \mu(x)=\int_{0}^{\infty} \delta(x-1/(2N)) F(N)dN.
\end{equation}
We expect that $\mu(x)$ will have some general properties, independent of the particular characteristics of $F(N)dN$. 
For instance, in many realistic scenarios $\mu(x)$ will be a function that is nearly zero for frequencies 
$x>x_\ast$, with $x_\ast=1/(2N_{min})$ a very small
characteristic frequency associated with the inverse of the minimum census population size.

Other phenomena that might not be properly captured by the simple mutational model in 
Eqs. \eqref{mutation1} and \eqref{mutation2}, consist of organisms with partially overlapping generations,
and organisms in which mutations in gametes arise from somatic mutations. When an organism 
has a mating pattern that violates the assumption
of non-overlapping generations (e.g. humans), the generation time in the model 
Eq. \eqref{WF} is interpreted as an \emph{average generation time}. 
Hence, during a generation unit, there is time enough for some individuals 
to be born with new mutations at the beginning of the generation time, and to reproduce themselves
by the end of a generation unit. This implies that after one average generation, there can exist
new identical mutations in more than one chromosome.
Similarly, when the gametes of an organism originate from somatic tissue, they inherit
de novo mutations that arose in the soma after multiple cell divisions. If the individuals of this organism can
have more than one offspring per generation, one expects to find the same new mutation, in the same site,
in more than one chromosome per generation.

Because of these different biological phenomena,
we believe that the notion of \emph{effective mutation density}, $\mu(x)$, 
is a more general way to describe mutations in natural populations. The \emph{effective mutation density}
describes the average frequency distribution of new mutations per generation, in one population, 
after taking into account the effects due to stochastic changes in census population size, 
non-overlapping generations and/or mutations of somatic origin.
From a numerical point of view, effective mutation densities are a useful tool to avoid
the numerical instabilities associated with polynomial expansions of non-smooth functions (e.g. Dirac deltas) that
appear in the standard approaches to mutation influx.
As we show later when we discuss the continuous limit of the infinite sites model, 
different effective mutation densities can yield predictions
which are identical to predictions of models based on Eq. \eqref{mutation2}. 

\subsection{More than one population}

Here, we show how to incorporate arbitrarily more populations, and migration flow between them.
Generally, for the state in the chain we consider a discrete 
random variable $\vec{X}$ which takes values in the $K$-dimensional lattice of derived allele frequencies:
\begin{equation}
\label{vector_mc}
\vec{X} = \left(
\begin{array}{c}
i_1/(2N_{1})\\
i_2/(2N_{2})\\
\vdots\\
i_K/(2N_{K})
\end{array}
\right),
\end{equation}
with $K$ the number of populations, and $0\leq i_a \leq 2N_{a}$. 
For simplicity, we use a single index notation, $0\leq I\leq \prod_a 2N_{a}$, 
to label the states where the random variable $\vec{X}$ takes values. 
The random variable $\vec{X}=I$ jumps to the state 
$\vec{X}=J$ at a discrete generation unit, with prescribed probability $P(J\vert I)$. 
The density of states in this multi-population setup is $f_I(t)$, and the difference equations that describe
its dynamics are equivalent to Eq. \eqref{chain_general}.
The transition matrix $\hat{P}=P(J\vert I)$ incorporates 
random drift and migration events between populations. New mutations enter each population with
an effective mutation vector $\vec{\mu}$
\begin{equation}
\label{mutat}
\vec{\mu} = \left(
\begin{array}{c}
0\\
\vdots\\
\mu^a_j\\
\vdots\\
0
\end{array}
\right);
\end{equation}
in the standard model, the mutation density is $\mu^a_j=2N_a \nu \delta_{1,j}$.

The Markov chain for a Wright-Fisher process for two independent populations is defined
by the transition matrix $\hat{P}_{j_1 j_2;i_1 i_2}=$
\begin{eqnarray}\label{WF2}
Bi(j_1; 2N_{e,1},i_1/(2N_{e,1}))Bi(j_2; 2N_{e,2},i_2/(2N_{e,2})),
\end{eqnarray}
where $Bi(j;k,p)$ stands for the binomial distribution with index $k$ and parameter $p$. 
Also, we can introduce  migration between populations, by sampling a constant
number of alleles $n_{ab}$ in population $a$ that become part of the allele pool in population $b$. Thus, in a model
of two populations with migration, one considers the transition matrix 
\begin{eqnarray}
\hat{P}_{j_1 j_2;i_1 i_2}=\sum_{k_1,k_2=0}^{k_1=n_{21},k_2=n_{12}} Bi(j_1-k_1; 2N_{e,1}-n_{21}, i_1/(2N_{e,1}))Bi(k_1; n_{21}, i_2/(2N_{e,2})) 
\times\nonumber \\
\times Bi(j_2-k_2; 2N_{e,2}-n_{12}, i_2/(2N_{e,2}))Bi(k_2; n_{12}, i_1/(2N_{e,1})).\label{WF3}
\end{eqnarray}
In this model the parameter space is given by the effective population sizes $N_{e,1}$ and $N_{e,2}$, and the scaled migration rates
$n_{21}$ and $n_{12}$.
This process is generalizable to an arbitrary number of populations in a straightforward way.

\section{Diffusion approximation}

Diffusion approximations to finite Markov chains have a 
distinguished history in population genetics, dating back to Wright and Fisher. This approach can be used to describe 
the time evolution of derived allele frequencies in several populations, with relatively large population sizes. 
This approximation applies when the population sizes $N_a$ are large (if $N_e>50$, the binomial distribution with index 
$2N_e$ can be accurately approximated by the Gaussian distribution used in the diffusion limit) and 
migration rates are of order $1/N_e$.

In the large population size limit, the state space spanned by vectors such as Eq. \eqref{vector_mc} converges to
the continuous space $[0,\, 1]^K$. The density of states $f_I(t)$ on the state space will converge
to a continuous density $\phi(x,t)$ on $[0,\, 1]^K$. The time evolution of $\phi(x,t)$ depends on how the inifinitesimal
change $\delta \vec{X}$,
$$
\vec{X}(t+\delta t)=\vec{X}(t)+\delta \vec{X},
$$
is distributed. If the mean of the $\delta \vec{X}$ distribution is $M(\vec{X},t)$
and the covariance matrix is $C(\vec{X},t)$, the time continuous limit $\delta t\to 0^+$ of the process $\vec{X}(t)$ 
is well defined. In the small, but finite, limit of $\delta t$ the stochastic process obeys the equation 
\begin{equation}\label{ito}
\vec{X}(t+\delta t) = \vec{X}(t) + M(\vec{X},t)\delta t + \sigma(\vec{X},t)\vec{\epsilon} \sqrt{\delta t},
\end{equation}
where $\vec{\epsilon}$ is a standard normal random variable 
(with zero mean and unit covariance matrix) in $\mathbb{R}^K$, $\sigma(\vec{X},t)$ is a square root of the covariance matrix 
$C(\vec{X},t)=\sigma \sigma^T (\vec{X},t)$, and $\delta t$ is a finite, but very small, time step.

In the diffusion approximation to the discrete Markov chain, the process is described as a time
continuous stochastic process governed by Gaussian jumps of prescribed variance and mean. 
These processes can be denoted using the notation of stochastic differential equations:
\begin{equation}\label{sde}
dX_t^a=M^a(X_t,t)dt + \sum_{b=1}^K \sigma^{ab}(X_t,t)dW^b_t,
\end{equation}
where $dW^b$ is the infinitesimal element of noise given by standard Brownian motion in $K$-dimensions, and
$\sigma$ is the square root matrix of the covariance matrix $C=\sigma \sigma^T$, \cite{shreve}. 
The diffusion generator associated with Eq. \eqref{sde} is
\begin{equation}\label{gen}
\mathcal{L}=\sum_{a=1}^K M^a(x,t)\frac{\partial}{\partial x^a} + \frac{1}{2}\sum_{b=1}^K C^{ab}(x,t)\frac{\partial^2}{\partial x^a \partial x^b}.
\end{equation}
Thus, if $\phi(x,t=0)$ is the density of allele frequencies at time $0$, the time evolution of $\phi(x,t)$ will be
governed by the forward Kolmogorov equation
\begin{equation}\label{kolmo}
\frac{\partial \phi(x,t)}{\partial t}=\sum_{a,b=1}^K
\frac{1}{2}\frac{\partial^2}{\partial x^a \partial x^b}\left[ C^{ab}(x,t) \phi(x,t) \right] - 
\sum_{a=1}^K \frac{\partial}{\partial x^a}\left[ M^a(x,t) \phi(x,t) \right] +\rho(x,t).
\end{equation}
Here, $\rho(x,t)$ is the continuous limit of $\mu_j$ in Eq. \eqref{chain_general}, 
that describes the net influx of polymorphisms in the population per generation.

\subsubsection{Modeling migration flow and random drift}

The continuous limit of the Markov chain defined in Eq. \eqref{WF3}, in the case of $K$ diploid populations and in
the weak migration limit, has as associated moments
\begin{eqnarray}
M^a(x,t)=\sum_b m_{ab}(x_b-x_a),\label{mean3}\\
C^{ab}(x,t)= \delta^{ab}\frac{x_a(1-x_a)}{2N_{e,a}} \label{migration_selection},
\end{eqnarray}
with $\delta^{ab}$ the Kronecker delta ($\delta^{ab}=1$ if $a=b$ and $\delta^{ab}=0$ otherwise).
The matrix element $m_{ab}=n_{ab}/(2N_a)$ defines the migration rate from the $b^{th}$ population 
to the $a^{th}$ population.

Thus, associated with this process one has the Kolmogorov forward equations 
\begin{eqnarray}
\frac{\partial}{\partial t} \phi(x,t)= \sum_{a,b}
\frac{1}{2}\frac{\partial^2}{\partial x_{a}\partial x_{b}}\left( \delta^{ab}\frac{x_a(1-x_a)}{2N_{e,a}} \phi(x,t)\right) 
\nonumber\\
-\frac{\partial}{\partial x_a}\left( m_{ab}(x_b-x_a) \phi(x,t)\right) +\rho(x,t)\label{migration_selection2}.
\end{eqnarray}
Eq. \eqref{migration_selection2} describes the time evolution of the frequency spectrum density under random drift 
and migration events between populations, given an initial density and absorbing boundary conditions (see below). 
The inhomogenous term $\rho(x,t)$ models the total incoming/outgoing flow of SNPs per generation into the $K$-cube which is not due to 
the diffusion flow, $j^a=-M^a \phi + \partial^b(C^{ab}\phi)$, at the boundary. This total flow depends on mutation events that
generate \emph{de novo} SNPs: inflow from higher dimensional components of the allele density (see below), 
inflow from migration events from lower dimensional components of the allele density, and the outflow of migration
events towards higher dimensional components.
If there is not a positive influx of SNPs, the density would converge to $\phi(x,t)\to 0$ as $t\to\infty$.
In order to understand the probability flow between different components of the density of alleles, we will have
to study how the boundary conditions are defined precisely.

\subsection{Boundary Conditions}

Understanding the boundary conditions in this problem is one of the most challenging tasks. 
In Kimura's famous solution to the problem of pure random drift in one population, \cite{kimuraI},
he required the solutions to the diffusion equation to be finite at the 
boundaries $x=0$ and $x=1$. This boundary condition is absorbing. 
The points $x=0$ and $x=1$ describe states where SNPs reach the fixation of
their ancestral or derived states.

\begin{figure}[hpbt]
\begin{center}
\vspace{0.5cm}
    \includegraphics[width=7.2cm]{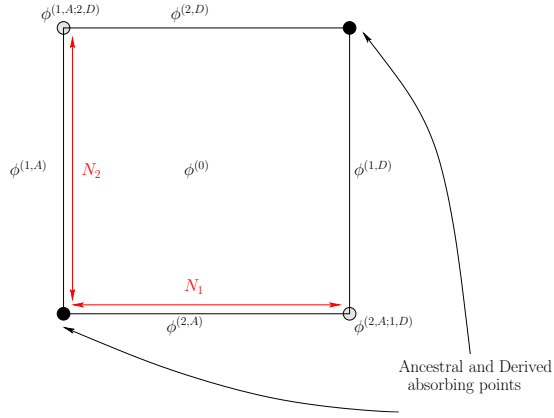}
\caption{\label{figboundary2}
Decomposition of the singular probability density, for two populations, on the two-dimensional 
bulk and the different subdimensional boundary components.
}
\end{center}
\end{figure}

\begin{figure}[hpbt]
\begin{center}
\vspace{0.5cm}
    \includegraphics[width=7.2cm]{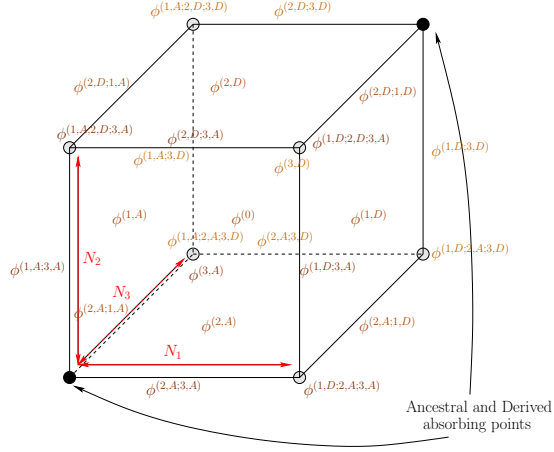}
\caption{\label{figboundary3}
Decomposition of the singular probability density, for three populations, on the three 
dimensional bulk and the different subdimensional boundary components.
}
\end{center}
\end{figure}

If we consider $K$ populations, the natural generalization of Kimura's boundary conditions can be derived
by studying the possible stochastic histories of single diallelic SNPs segregating in the $K$ populations. 
A SNP which is initially polymorphic in all the $K$ populations can reach the fixation of its derived or ancestral state
in one population while still being polymorphic in the remaining $K-1$ populations. More generally, a SNP can
be polymorphic in $K-\alpha$ populations, while its state can be fixated in the remaining 
$\alpha$ populations. A convenient way of visualizing this is to look at the geometry of the $K$-cube of allele
frequencies, and the different subdimensional components of its boundary 
(see examples Fig. \ref{figboundary2} and Fig. \ref{figboundary3} for the 2-cube and 3-cube).
A $K$-cube's boundary can be decomposed as a set of cubes of lower dimensionality, from $(K-1)$-cubes up to $0$-cubes or
points. The number of boundary components of codimension $\alpha$, i.e. the number of $(K-\alpha)$-cubes, contained in the boundary of the $K$-cube is
\begin{equation}
\label{boundary_components}
\#\, (K-\alpha)\,\mathrm{cubes}\,= \frac{2^\alpha K!}{(K-\alpha)! \alpha!}.
\end{equation}

The most important set of boundary components are the $(K-1)$-cubes, because any other boundary component
can be expressed as the intersection of a finite number of $(K-1)$-cubes at the boundary. We identify 
each $2K$ codimension-one boundary component by the population where the SNPs are not polymorphic, and by 
the state that is fixated in this population (Derived or Ancestral). For example, the component
$(i,A)$ is defined as the set of points in the $K$-cube that obeys the equation $x_i=0$, and the component
$(i,D)$ is defined by the equation $x_i=1$. Therefore, any codimension $\alpha$ boundary component can
be expressed as the intersection
\begin{equation}
\label{boundary_component}
(i_1\, S_1)\cap (i_2\, S_2) \cap \cdots (i_\alpha\, S_\alpha) = \{ x\in [0,1]^K \vert x_{i_1}=\delta_{S_1, D};\, x_{i_2}=\delta_{S_2, D};\ldots
; x_{i_\alpha}=\delta_{S_\alpha, D} \},
\end{equation}
with $i_\alpha\neq i_\beta$ when $\alpha\neq \beta$, $\delta_{S,D}=1$ for the derived state $S=D$, and $\delta_{S,D}=0$ 
for the ancestral state $S=A$.

To each boundary component of codimension $\alpha$ we associate a $(K-\alpha)$-dimensional density of derived allele
frequencies that are polymorphic only on the corresponding $K-\alpha$ populations, while are fixated in the other
$\alpha$ populations. In this way, $\phi^{(0)}$ denotes the bulk probability density, $\{ \phi^{(i,S_i)} \}_{i=1}^{i=K}$ (with the state
$S_i$ being either ancestral $S_i=A$ or derived $S_i=D$) are the 
$2K$ codimension-one densities, $\{ \phi^{(i,S_i; j,S_j)} \}_{i\neq j}$ the codimension 2 densities, etc. This
decomposition is illustrated in the case of 2 and 3 populations in Fig. \ref{figboundary3} and
Fig. \ref{figboundary2}.

In this notation, we write the density of derived alleles segregating on $K$ populations as the generalized
probability function
\begin{eqnarray}
\phi(x,t)=\phi^{(0)}(x,t) + \sum_{i=1}^K \left( \phi^{(i,A)}(x,t)\delta(x_i) + 
\phi^{(i,D)}(x,t)\delta(x_i-1)\right) + \nonumber \\
\sum_{i=1, j\neq i}^K \Big( \phi^{(i,A;j,A)}(x,t)\delta(x_i)\delta(x_j) +  \phi^{(i,A;j,D)}(x,t)\delta(x_i)\delta(x_j-1)
\nonumber\\
+ \phi^{(i,D; j,D)}(x,t)\delta(x_i-1)\delta(x_j-1)\Big) +  \sum_{i=1, j\neq i, k\neq i, k\neq j}^K\ldots
\label{total_density}
\end{eqnarray}
with $\delta(\cdot)$ the Dirac delta-function.
The points $(1,\, A)\cap (2,\, A) \cap \cdots (K,\, A)$ and $(1,\, D)\cap (2,\, D) \cap \cdots (K,\, D)$ are the 
universal fixation boundaries, and they do not contribute to the total density of alleles in Eq. \eqref{total_density}.
It is useful to write the probability density $\phi(x,t)$ using such decomposition, because despite being a 
singular generalized function, each boundary component $\phi^{(i,S_1;j,S_2,\ldots )}(x,t)$ will be,
most of the time, a regular analytic function.

The dynamics of the boundary components $\phi^{(i_1,S_1;i_2,S_2,\ldots )}(x,t)$ are governed by diffusion equations,
with an inhomogenous term, of the type
\begin{eqnarray}
\frac{\partial}{\partial t} \phi^{(i_1,S_1;i_2,S_2,\ldots )}(x,t)= \sum_{a,b\neq i_1,i_2\ldots}
\frac{1}{2}\frac{\partial^2}{\partial x_{a}\partial x_{b}}\left( \delta^{ab}\frac{x_a(1-x_a)}{2N_{e,a}} 
\phi^{(i_1,S_1;i_2,S_2,\ldots )}(x,t)\right) 
\nonumber\\
-\frac{\partial}{\partial x_a}\left( m_{ab}(x_b-x_a) \phi^{(i_1,S_1;i_2,S_2,\ldots )}(x,t)\right) 
+\rho^{(i_1,S_1;i_2,S_2,\ldots )}(x,t)\label{migration_selection33},
\end{eqnarray}
with $\rho^{(i_1,S_1;i_2,S_2,\ldots )}(x,t)$ the net incoming/outgoing flow into the boundary component
$(i_1,S_1)$ $\cap$ $(i_2,S_2)$ $\cap\ldots$. The $\rho$ term can be interpreted as an interaction term 
between different boundary components.

More precisely, $\rho^{(i_1,S_1;i_2,S_2;\ldots )}(x,t)$ consists of four terms
\begin{equation}
\rho^{(i_1,S_1;i_2,S_2;\ldots; i_\alpha,S_\alpha )}(x,t)= \rho_{mut}(x,t) + \rho_{drift}(x,t) +
\rho_{in_m}(x,t) + \rho_{out_m}(x,t).
\end{equation}
$\rho_{mut}(x,t)$ is the influx of spontaneous mutations (only present in codimension $K-1$), 
$\rho_{drift}(x,t)$ consists of the boundary inflow from codimension $\alpha-1$ components that have 
$(i_1,S_1)$ $\cap$ $(i_2,S_2)$ $\cap\ldots \cap (i_\alpha,S_\alpha)$ as a boundary component, $\rho_{in_m}(x,t)$ represents the incoming flow
due to migration events from lower dimensional boundary components, and
$\rho_{out_m}(x,t)$ is the outflow due to migration events from $(i_1,S_1)$ $\cap$ $(i_2,S_2)$ $\cap\ldots$ 
$(i_\alpha,S_\alpha)$ towards higher dimensional components that have $(i_1,S_1)$ $\cap$ $(i_2,S_2)$ $\cap\ldots$ 
$(i_\alpha,S_\alpha)$ as a boundary component.

We can write in a more precise way each term in $\rho^{(i_1,S_1;i_2,S_2,\ldots )}(x,t)$, as follows:
\begin{itemize}
\item $\rho_{mut}$:
\begin{equation}
\rho^{(i_1,S_1;i_2,S_2;\ldots; i_\alpha,S_\alpha )}_{mut}(x,t)=\sum_a 2N_{e,a} u \delta_{\alpha,K-1}
\mu(x_a) \prod_{j=1}^\alpha (1-\delta_{i_j,a}),
\label{mutation}
\end{equation}
with $\delta_{\alpha,K-1}=1$ if $\alpha=K-1$, $\delta_{\alpha,K-1}=0$ if $\alpha\neq K-1$, and $\mu(x_a)$ is the mutation density 
(in the classical theory, $\mu(x_a)=\delta(x_a-1/(2N_{e,a}))$). 
\item $\rho_{drift}$: Assuming that the first derivatives of $\phi(x,t)$ are finite at the boundary,
\begin{eqnarray}
\rho^{(i_1,S_1;i_2,S_2;\ldots; i_\alpha,S_\alpha )}_{drift}(x,t)=\sum_{j_{\alpha}=i_1}^{i_\alpha} \sum_a \Big(\delta_{S_{j_\alpha},A} \Big(\frac{1}{4N_{e,a}}
 -\sum_{b}m_{ab}x_b \Big)\label{drift}\\ +\delta_{S_{j_\alpha},D} \Big(\frac{1}{4N_{e,a}}
 -\sum_{b}m_{ab}(1-x_b) \Big) \Big)\phi^{(j_1,S_1;j_2,S_2;\ldots; j_{\alpha-1},S_{\alpha-1} )}(x_{j_{\alpha}}=\delta_{S_{j_\alpha},D}, x,t),\nonumber
\end{eqnarray}
where the sum over $j_{\alpha}$ is over all components of codimension $\alpha-1$ that have 
$(i_1,S_1)$ $\cap$ $(i_2,S_2)$ $\cap\ldots$ $(i_\alpha,S_\alpha)$ as a boundary component. $\delta_{S_{j_\alpha},A}$
is $1$ when $S_{j_\alpha}=A$ and $0$ when $S_{j_\alpha}=D$; similarly $\delta_{S_{j_\alpha},D}$ is one when
$S_{j_\alpha}=D$ and zero when $S_{j_\alpha}=A$. The sum over $a$ and $b$
is over all populations that are not $j_1,j_2,\ldots$ $j_{\alpha}$.
\item $\rho_{in_m}$: 
Here, and throughout, $c_\alpha$ is a shorthand for the boundary component 
$(i_1,S_1;$ $i_2,S_2;$ $\ldots;$ $i_\alpha,S_\alpha )$.
$\rho^{(c_\alpha)}_{in_m}(x,t)$ represents the total incoming flow due 
to migration events of SNPs that are contained in densities of SNPs 
located at boundary components of $c_\alpha$.
If $\mathcal{B}_d(c_\alpha)$
is the set of boundary components of $c_\alpha$ with fixed codimension $d$ ($\alpha<d\leq K$),
\begin{eqnarray}
\mathcal{B}_d=\{
(i_1,S_1;i_2,S_2;\ldots;i_\alpha,S_\alpha;j_{\alpha+1},S_{\alpha+1};\ldots j_{d},S_{d})
\}_{j_{\alpha+1},\ldots,j_{d}},\nonumber
\end{eqnarray}
then $\rho^{(c_\alpha)}_{in_m}(x,t)$ can be written as the sum of contributions from all boundary components 
in $\mathcal{B}_d$,  for all codimensions $d=\alpha+1, \alpha+2,\ldots, K$, and for all possible migration events 
from elements $q$ in $\mathcal{B}_d(c_\alpha)$ to $c_\alpha$:
\begin{equation}
\label{incoming_migrants}
\quad\quad \rho^{(bc_\alpha)}_{in_m}(x,t)=
\sum_{d=\alpha+1}^K \sum_{q\in\mathcal{B}_d} \phi^{(q)}(x,t) \times \left( \sum_{e\in \Gamma(q\to c_\alpha)} p(e)
\prod_{k=\alpha+1}^d \delta(x_{i_k}-f^{e}_{i_k} ) \right).
\end{equation}
Here, $\Gamma(q\to c_\alpha)$ is the set of all possible migrations events from SNPs in $\phi^{(q)}$ to
$\phi^{(c_\alpha)}$, $p(e)$ denotes the probability of occurence of the migration event $e$, and 
$0<f^{e}_i\in c_\alpha$ denotes the expected frequency, in the $i^{th}$-population, of a SNP that enters 
$c_\alpha$ after the event $e$. We provide below a more precise description of Eq. \eqref{incoming_migrants},
such as a description of the event space 
$\Gamma(q\to c_\alpha)$, the corresponding probabilities of occurence and expected frequencies.

\item $\rho_{out_m}$: Denotes the outflow of SNPs due to migration events to higher dimensional boundary
components. In other words, $\rho_{out_m}^{(c_\alpha)}$ measures the rate of loss of SNPs in 
$\phi^{(c_\alpha)}$, because of migration flow towards boundary components 
of codimension $d<\alpha$, that have $c_\alpha$ as a boundary component. Let 
$\mathbf{I}_{\partial q, c_\alpha}$ be a discrete function that returns 1 when $c_\alpha$ is a boundary
component of $q$, and zero when it is not. Thus,
\begin{eqnarray}
\rho^{(c_\alpha)}_{out_m}(x,t)= -
\phi^{(c_\alpha)}(x,t) \times  
\sum_{d=0}^{\alpha-1} \sum_{q\in\mathcal{B}_d} \mathbf{I}_{\partial q, c_\alpha}
\left( \sum_{e\in \Gamma(c_\alpha\to q)} p(e)\right). \label{outgoing_migrants}
\end{eqnarray}
\end{itemize}

To compute Eq. \eqref{incoming_migrants} and Eq. \eqref{outgoing_migrants}, the use of approximations is unavoidable. 
In principle, one could use the transition probabilities of the finite 
Markov chain to estimate the probabilities of different migration events and their expected allele frequencies. 
However, there is a simpler approximation, 
which is consistent with the weak migration limit in which the diffusion equation is derived. 

This approximation follows from the observation that at the boundary $x_a=\delta_{S,D}$, the strength of 
random drift along the population $a$ vanishes ($x_a(1-x_a)=0$), and hence, the infinitesimal change in $x_a$ obeys
a deterministic equation:
\begin{equation}
\label{freq_a}
\frac{dx_a}{dt}=\sum_b \delta_{S,A}m_{ab}x_b -\delta_{S,D}m_{ab}(1-x_b).
\end{equation}
Eq. \eqref{freq_a} implies that a migration event from several populations $b$, to a target 
population $a$, can push the frequency $x_a$ of a SNP out of the boundary where it was initially fixated 
($x_a=\delta_{S,D}$). 

Therefore, given a $K$-cube, a boundary component $c_\alpha$ (of codimension $\alpha$), and a 
boundary component $q$ (of codimension $\beta>\alpha$) of $c_\alpha$, we say
that there will exist migration flow from $q$ to $c_\alpha$, if and only if
\begin{eqnarray}
\frac{dx_{a_t}}{dt}= \sum_{b=1}^K \delta_{S_{a_t},A}m_{a_t b}x_{b} -\delta_{S_{a_t},D}m_{a_t b}(1-x_{b})\neq 0,
\label{freq_b}\\
\frac{dx_{n}}{dt}= \sum_{b=1}^K \delta_{S_{n},A}m_{n b}x_{b} -\delta_{S_{n},D}m_{n b}(1-x_{b})=0,
\label{freq_c}
\end{eqnarray}
where $\{x_n \}_{n=1}^{\alpha}$ denote the allele frequencies of SNPs 
which are fixated at the boundary components $c_\alpha$ and $q$, $\{ a_t \}_{a_t=\alpha+1}^{K}$ are
the populations at $c_\alpha$ whose allele frequencies are polymorphic, and the frequencies $x_b$ are defined
at the boundary component $q$ (which means that $x_b$ is polymorphic as long as $b>\beta$, and is $\delta_{S_b, D}$ 
otherwise). It is important to realize that $x_{b}$ can be $0$ or $1$ at $q$, and a migration event to $c_\alpha$
can still bring alleles of the opposite state that is fixated in the target population.

In this approximation, $\Gamma(q\to c_\alpha)$ consists of a single element, 
and $p(e)$ can be zero or one. 
If Eq. \eqref{freq_b} and Eq. \eqref{freq_c} are satisfied, the migration event in $\Gamma(q\to c_\alpha)$ has probability $p(e)=1$, and
the expected frequencies are
\begin{equation}
f_{a_t}= \sum_{b=1}^K \delta_{S_{a_t},A}m_{a_t b}x_{b} -\delta_{S_{a_t},D}m_{a_t b}(1-x_{b}).
\end{equation}
If Eq. \eqref{freq_b} and Eq. \eqref{freq_c} are not satisfied, $p(e)=0$, and we say that there is not migration flow from $q$ to $c_\alpha$.

\subsection{Effective Mutation Densities}

Given a constant spontaneous mutation rate in the species under study, of $u$ ``base substitutions per site and per generation,''
and expected number of sites $\nu=L\times u$ where new mutations appear in the population each generation,  
the total number of \emph{de novo} mutant sites that appear in the population $a$, every generation, is $2N_{e,a} \nu$.
We can model this constant influx of mutations by adding a Dirac delta term
\begin{equation}
2N_{e,a} \nu \delta(x_a-1/(2N_{e,a})),
\end{equation}
to the $K$ diffusion equations that govern the ancestral components of codimension $K-1$ 
$\phi^{(i_1\, A)\cap (i_2\, A) \cap \cdots (i_{K-1}\, A)}(x,t)$. However, as we discussed above, more generally we work
with an effective mutation density
\begin{equation}
2N_{e,a} \nu \mu(x_a).
\end{equation}

\begin{figure}[hpbt]
\begin{center}
\vspace{0.5cm}
    \includegraphics[width=7.5cm]{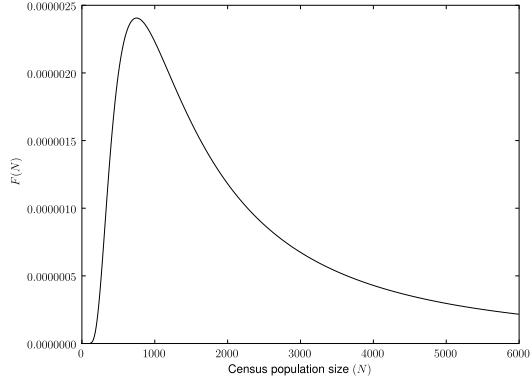}
\caption{\label{mutation_density} A model for a stochastic census population size, with exponential decay
in the small population size limit, a quadratic decay $\sim N^{-2}$ in the large population size limit,
and a population peak at $N=1000$.
}
\end{center}
\end{figure}

As a particular example of an effective mutation density, we consider a 
stochastic census population size, which is a random variable distributed as 
\begin{equation}
F(N)dN = \frac{c}{2 N^2}\exp(-\kappa/(2N))dN.
\end{equation}
This distribution avoids extremely small populations by an exponential tail, while large population
sizes are distributed as $\sim N^{-2}$, as shown in Fig. \ref{mutation_density}.
In this model, we keep constant the effective population size $N_e$ that defines the variance of random drift in
Eq. \eqref{migration_selection}.
Thus, the mutation density will be
\begin{equation}
\label{mutation_density_gaussian}
\mu(x)=\int_{0}^\infty \delta(x-1/2M) \frac{c}{2 M^2}\exp(-\kappa/2M)dM.
\end{equation}
We can integrate Eq. \eqref{mutation_density_gaussian} exactly, by making the change of variables $y=1/2M$,
$dM=-dy/2y^2$:
\begin{eqnarray}
\mu(x)=c\exp(-\kappa x). \label{mutation_density_gaussianII}
\end{eqnarray}

\subsection{Population splitting events}

So far we have studied how the allele frequency density changes as a function of time while the number of populations $K$ remains constant.
When two populations split, the diffusion jumps to dimension $K+1$, and the corresponding density will obey the 
time evolution defined by Eq. \eqref{migration_selection2} for $K+1$ populations, with different population sizes and migration parameters. 
The initial
density $\phi_{K+1}(x,x_{K+1},t)$ in the $K+1$ diffusion problem is determined from the density $\phi_{K}(x,t)$, before the populations split. 
Therefore, if population $K+1$ was formed by $N_{f,a}$ migrant founders from the $a^{th}$ population, then
\begin{equation}
\label{founder}
\phi_{K+1}(x, x_{K+1},t) = \phi_{K}(x,t) \sqrt{\frac{N_{f,a}}{\pi x_a(1-x_a)}} e^{-N_{f,a}(x_a-x_{K+1})^2/(x_{a}(1-x_{a}))}.
\end{equation}
This formula is derived by considering the binomial sampling of $2N_{f,a}$ chromosomes from population $a$, and using 
the Gaussian approximation for the binomial distribution with $2N_{f,a}$ degrees of freedom and parameter $x_a$. 
In the limit $N_{f,a}\to\infty$, Eq. \eqref{founder} simplifies to
\begin{equation}
\label{founderII}
\phi_{K+1}(x, x_{K+1},t) = \phi_{K}(x,t) \delta(x_a-x_{K+1}),
\end{equation}
with $\delta(x)$ the Dirac delta. Additionally, if the new population is formed by migrants from two populations merging,
with a proportion $f$ from population $i$ and $1-f$ from population $b$, then
\begin{equation}
\label{founderIII}
\phi_{K+1}(x, x_{K+1},t) = \phi_{K}(x,t) \delta(fx_a+(1-f)x_b-x_{K+1}).
\end{equation}
In the diffusion framework, one can also deal with populations that go extinct or become completely isolated.
More precisely, if we remove the $a^{th}$ population, the initial density in the $K-1$ dimensional problem will be
\begin{equation}
\label{extinction}
\phi_{K-1}(\tilde{x},t) = \int_{[0,1]} \phi_{K}(x,t) dx_{a},
\end{equation}
with $\tilde{x}$ denoting the vector $\tilde{x}=(x_1,x_2,\ldots, x_{a-1}, x_{a+1}, \ldots x_K)$.

\section{Solution to the diffusion equations using spectral methods}

The idea behind spectral methods consists of borrowing analytical methods from the theory of Hilbert spaces to transform 
a partial differential equation, such as Eq. \eqref{migration_selection2}, 
into an ordinary differential equation that can be integrated numerically using, for instance, a Runge-Kutta method.

In general, the problems in which we are interested are mixed initial-boundary value problems of the form
\begin{eqnarray}
\frac{\partial \phi(x,t)}{\partial t} = L_{FP}(x,t)\phi(x,t) +\rho(x,t),\quad x\in D=[0,\, 1]^K \label{fp},\\
B(x)\phi(x,t)=0,\quad x\in\partial D, t>0,\label{fpb}\\
\phi(x,0)=g(x),\quad x\in D,\label{boundary}
\end{eqnarray}
where $D=[0,\, 1]^K$ is the frequency spectrum domain with boundary $\partial D$, $L_{FP}(x,t)$ is a linear differential 
operator also known as the Fokker-Planck operator, $\rho(x,t)$ is a function, and $B(x)$ is the linear boundary operator that 
defines the boundary condition.  In this paper, we are interested in the particular set of PDEs defined in 
Eq. \eqref{migration_selection2}, although we sometimes keep the notation of Eq. \eqref{fp} as a shorthand.

We assume that $\phi(x,t)$ is for all $t$ an element of a Hilbert space $\mathcal{H}$ of square integrable functions,
and associated $L^2$-product $\langle\, ,\,\rangle_{L^2}$. Furthermore, we assume that all functions in $\mathcal{H}$ satisfy the boundary conditions 
imposed by Eq. \eqref{fpb}. In spectral methods we consider a complete orthogonal basis of functions for 
$\mathcal{H}$, that we denote by $\{ \psi_i(x)\}_{i=0}^{\infty} $, which obeys
\begin{equation}\label{orth1}
\langle \psi_i ,\, \psi_j\rangle_{L^2}=h_{i}\delta_{ij},
\end{equation}
with $h_i$ a function of $i$ that depends on the particular choice of basis functions.
One then approximates $\phi(x,t)$ as the truncated expansion
\begin{equation}\label{proj}
\mathcal{P}_{\Lambda}\phi(x,t)= \sum_{i=0}^{\Lambda-1} \alpha_i(t) \psi_i (x).
\end{equation}
Similarly, one approximates the PDE in Eq. \eqref{fp} by projecting it onto the finite dimensional basis 
$\{ \psi_i(x)\}_{i=0}^{\Lambda-1} $, as
\begin{equation}
\label{ode1}
\frac{\partial }{\partial t} \mathcal{P}_{\Lambda} \phi(x,t) = \mathcal{P}_{\Lambda} L_{FP}(x,t)\mathcal{P}_{\Lambda}\phi(x,t) 
+\mathcal{P}_{\Lambda}\rho(x,t).
\end{equation}
By $\mathcal{H}_{\Lambda}$ we denote the finite dimensional space spanned by $\{ \psi_i(x)\}_{i=0}^{\Lambda-1}$, and by
$\mathcal{P}_\Lambda$ the corresponding projector $\mathcal{H}\to \mathcal{H}_{\Lambda}$. If 
\begin{eqnarray}
\mathcal{P}_{\Lambda} L_{FP}(x,t)\mathcal{P}_{\Lambda}\phi(x,t)=\sum_{i,j=0}^{\Lambda-1}\omega_{ij}(t) \psi_i(x) \alpha_j(t) \nonumber\\ 
\mathcal{P}_{\Lambda}\rho(x,t) =\sum_{i=0}^{\Lambda-1} \beta_i \psi_i(x) \nonumber,
\end{eqnarray}
we can re-write the ODE in Eq. \eqref{ode1} using just \emph{modal} variables as
\begin{equation}
\label{ode2}
\frac{\partial \alpha_{i}(t) }{\partial t}  =  \sum_{j=0}^{\Lambda-1}\omega_{ij}(t)\alpha_{j}(t)+\beta_i.
\end{equation}
One can solve Eq. \eqref{ode2} by discretizing the time variable $t$, and using a standard numerical method
to integrate ODEs.
Therefore, the spectral solution to the diffusion PDE is expressed in the form of a truncated expansion, like Eq. \eqref{proj}, 
and has coefficients determined by the integral of Eq. \eqref{ode2}.

There are many different ways to construct sequences of approximating spaces $\mathcal{H}_\Lambda$ 
that converge to $\mathcal{H}$ in the limit $\Lambda\to\infty$, when the domain is the $K$-cube. Here, we follow other authors'
preferred choice \cite{spectral}, and choose
the basis of Chebyshev polynomials of the first kind. In the following section we introduce Chebyshev expansions 
and show why they are a preferred choice.

\subsection{Approximation of functions by Chebyshev expansions}

Let $\{ T_i(x) \}_{i=0}^\infty$ be the basis of Chebyshev polynomials of the first kind. They are the set of eigenfunctions
that solve the singular Sturm-Liouville problem
\begin{equation}\label{sturm_liouville}
\frac{d}{dx}\left( \sqrt{1-x^2}\frac{dT_i (x)}{dx} \right)+\frac{i^2}{\sqrt{1-x^2}}T_i (x)=0,
\end{equation}
with $i=0,1,\ldots,\infty$, and $-1\leq x\leq 1$. $\{ T_i(x) \}_{n=0}^\infty$ are orthogonal under the $L^2$-product
with weight function $w(x)=1/\sqrt{1-x^2}$:
\begin{equation}\label{l2prod}
\int_{[-1,1]} T_i(x) T_j(x)  \frac{dx}{\sqrt{1-x^2}} = \frac{\pi c_i}{2} \delta_{ij},
\end{equation}
where $c_0=2$ and $c_{i>0}=1$. 
This basis of polynomials is a natural basis for the approximation of functions on a finite interval
because the associated Gauss-Chebyshev quadrature formulae have an exact
closed form, the evaluation of the polynomials is very efficient, and the 
convergence properties of the Chebyshev expansions are excellent \cite{spectral}.

The Chebyshev polynomials of the first kind can be evaluated by using trigonometric functions, because of
the identity $T_i(x)=\cos(i\,\arccos(x))$. The derivatives of the basis functions can be computed by utilizing
the recursion 
\begin{equation}\label{recur}
T_i(x)= -\frac{1}{2(i-1)}T_i^\prime (x) + \frac{1}{2(i+1)}T_i^\prime (x),
\end{equation}
to express the derivative as
\begin{equation}\label{recur2}
T_i^\prime (x) = \sum_{j=0\,\vert j+i\,\, odd}^{i-1}\frac{1}{c_j} T_j(x).
\end{equation}
Similar formulae can be found for higher derivatives.
The coefficients in the expansion
\begin{equation}\label{approx}
\mathcal{P}_{\Lambda}f(x)=\sum_{i=0}^{\Lambda-1} a_i T_i(x),
\end{equation}
can be calculated by using the orthogonality relations of the basis functions 
\begin{equation}\label{coeffs1}
a_i=\frac{2}{\pi c_i} \int_{-1}^1 f(x) T_i(x)\frac{dx}{\sqrt{1-x^2}}.
\end{equation}
However, a direct evaluation of the continuous inner product, Eq. \eqref{coeffs1}, can be a source
of considerable problems, as in the case of the Fourier series. The classical solution lies in the 
introduction of a Gauss quadrature of the form
\begin{equation}\label{FT}
\frac{2}{\pi c_i} \int_{-1}^1 f(x) T_i(x)\frac{dx}{\sqrt{1-x^2}} \simeq \frac{2}{c_i Q}\sum_{k=1}^Q
f(x_k) T_i(x_k),\quad\quad x_k=\cos\left( \frac{2k-1}{2Q}\pi \right).
\end{equation}
If $f(x)$ is smooth enough, the finite sum
over $Q$ grid points in Eq. \eqref{FT} will converge quicker than $O(Q^{-1})$ to the exact formula \cite{numericalrecipes}.
As Eq. \eqref{FT} is equivalent to a discrete Fourier cosine transform, general results on the convergence
of cosine transforms apply also to this problem. One can see this relationship by considering
the change of variables $x=\cos y$:
\begin{equation}\label{FT2}
\frac{2}{\pi c_i} \int_{-1}^1 f(x) T_i(x)\frac{dx}{\sqrt{1-x^2}} = \frac{2}{\pi c_i} \int_{0}^\pi f(\cos y) \cos (iy)dy,
\end{equation}
and choosing $Q$ equally spaced abscissas in the interval $0\leq y\leq \pi$,
\begin{equation}\label{FT3}
\frac{2}{\pi c_i} \int_{0}^{\pi} f(\cos y) \cos (iy)dy\simeq \frac{2}{\pi c_i} \frac{\pi}{Q} \sum_{k=1}^Q
f\left[\cos\left( \frac{2k-1}{2Q}\pi\right) \right] \cos \left( i\frac{2k-1}{2Q}\pi \right).
\end{equation}

In order to study the convergence properties of the Chebyshev expansions Eq. \eqref{approx}, we exploit the rich
analytical structure of the Chebyshev polynomials. By using the identity Eq. \eqref{sturm_liouville}, one can 
re-write Eq. \eqref{coeffs1} as
\begin{equation}\label{conve}
a_i=-\frac{2}{\pi c_i i^2} \int_{-1}^1 f(x)\frac{d}{dx}\left[ \sqrt{1-x^2} \frac{d T_i(x)}{dx} \right] dx.
\end{equation}
If $f(x)$ is $C^1([-1,1])$ (i.e., if its first derivative is continuous), we can twice integrate by parts 
Eq. \eqref{conve} to obtain
\begin{equation}\label{conve2}
a_i=-\frac{2}{\pi c_i i^2} \int_{-1}^1 \sqrt{1-x^2}\frac{d}{dx}\left[ \sqrt{1-x^2} \frac{d f(x)}{dx} \right]  \frac{T_i(x)}{\sqrt{1-x^2}}
dx.
\end{equation}
We can repeat this process as many times as $f(x)$ is differentiable; thus, if $f(x)\in C^{2q-1}([-1,1])$
then 
\begin{equation}\label{conve3}
a_i=-\frac{2}{\pi c_i i^{2q}} \int_{-1}^1 \left[  \left(\sqrt{1-x^2}\frac{d}{dx} \sqrt{1-x^2} \frac{d}{dx}\right)^q f(x) \right]  T_i(x)
\frac{dx}{\sqrt{1-x^2}}.
\end{equation}
If we use the truncation error
\begin{equation}\label{trunca}
\Vert f(x)- \mathcal{P}_{\Lambda}f(x) \Vert_{L^2} = \left[ \int_{-1}^1 \left\vert f(x) - 
\sum_{i=0}^{\Lambda-1} a_i T_i(x) \right\vert^2 \frac{dx}{\sqrt{1-x^2}} \right]^{1/2}= \left( \sum_{i=\Lambda}^{\infty} 
\vert a_i\vert^2 \right)^{1/2}
\end{equation}
as a measure of convergence of the Chebyshev expansion, we may estimate its asymptotic expansion by calculating the rate 
of decrease of $a_i$. But as we showed in Eq. \eqref{conve3}, $\vert a_i\vert = \frac{c(q)}{i^{2q}}$, for some constant $c(q)$
if $f(x)\in C^{2q-1}([-1,1])$. Therefore, for large $\Lambda$ the error decreases as a power law
\begin{equation}\label{trunca2}
\Vert f(x)- \mathcal{P}_{\Lambda}f(x) \Vert_{L^2} = \left( \sum_{i=\Lambda}^{\infty} 
\vert a_i\vert^2 \right)^{1/2} \leq \frac{c}{\Lambda^{2q-1}},
\end{equation}
and if the function is infinitely differentiable ($q=\infty$), the corresponding Chebyshev series expansion will converge faster
than any power of $1/\Lambda$.

In the applications of this paper we will work with re-scaled Chebyshev polynomials.
As the Allele Frequency Spectrum is defined on the interval $[0,1]$, or direct products of it, 
we re-scale the Chebyshev polynomials to obtain an orthonormal basis on 
$[0,1]$. More precisely, the basis that we use is
$\{ R_i(x)=\alpha_iT_i((1-x)/2) \}_{i=0}^\infty$ with $x\in [0,1]$, $\alpha_0=1/\sqrt{\pi}$, $\alpha_{i>0}=\sqrt{2}/\sqrt{\pi}$,
$L^2$-product:
\begin{equation}\label{inner2}
\langle f, \, g \rangle = \int_{[0,1]}f(x)g(x)\frac{dx}{\sqrt{x(1-x)}},
\end{equation}
and orthonormality relations,
\begin{equation}\label{inner3}
\int_{[0,1]}R_i(x)R_j(x)\frac{dx}{\sqrt{x(1-x)}} = \delta_{ij}.
\end{equation}

\subsubsection{High-dimensional domains and spectral approximations of functional spaces}

The joint site frequency spectrum of $K$ populations can be defined as a density on $[0,1]^K$.
A natural basis of functions on the Hilbert space $L^2_w ([0,1]^K)$, comes from the tensor product of
one dimensional functions. More particularly, we consider the tensor product of Chebyshev polynomials
\begin{equation}\label{multipop}
\psi_{i_1,i_2,\ldots i_K}(x)=R_{i_1}(x_1)R_{i_2}(x_2)\cdots R_{i_K}(x_K),
\end{equation}
because $L^2_w ([0,1]^K)=L^2_w([0,1])\otimes \cdots \otimes L^2_w([0,1])$. Therefore, any square integrable function $F(x)$
under the $L^2$-product
\begin{equation}
\langle F(x),\, G(x)\rangle_w =\int_{[0,1]^K} F(x)G(x)\prod_{a=1}^K\frac{dx_a}{\sqrt{x_a(1-x_a)}},
\end{equation}
can be approximated as multi-dimensional Chebyshev expansion
\begin{equation} \label{expa1}
F(x)=\sum_{i_1=0}^{\Lambda_1 -1}\sum_{i_2=0}^{\Lambda_2 -1}\cdots \sum_{i_K=0}^{\Lambda_K -1} 
\alpha_{i_1,i_2,\ldots i_K} R_{i_1}(x_1)R_{i_2}(x_2)\cdots R_{i_K}(x_K).
\end{equation}
The truncation parameters $\Lambda_1$, $\Lambda_2\ldots$ can be fixed depending on the analytical properties 
of the set of functions that one wants to approximate and their corresponding truncation errors. There always exists
a trade-off between the accuracy of the approximation (the higher the $\Lambda$ the more accurate the approximation) 
and the speed of the implementation of the algorithm (the lower the $\Lambda$, the faster the algorithm); therefore,
choosing different values of $\Lambda_i$ will yield more optimal implementations of the algorithm.
Here, for simplicity in the notation, 
we use a unique truncation parameter $\Lambda=\Lambda_1=\cdots =\Lambda_K$. 

\subsection{Diffusion Operators in Modal Variables}

In order to approximate the PDEs defined in Eq. \eqref{migration_selection2} by a system of ODEs in the modal Chebyshev variables such 
as Eq. \eqref{ode2}, we need to project the Fokker-Planck operator in the Chebyshev basis spanned
by Eq. \eqref{multipop}. Later on we will show how to deal with the influx of mutations specified by the Dirac
deltas.

A direct projection of the Fokker-Planck operator onto the Chebyshev basis spanned by Eq. \eqref{multipop},
would require storing the coefficients in a huge matrix with $\Lambda^{2K}$ matrix elements. Fortunately,
the Fokker-Planck operator in our problem is very simple, and its non-trivial information can be stored in
just four sparse $\Lambda\times\Lambda$ matrices.
First, we need the random drift matrix
\begin{equation}
\label{RD}
D_{ij}=\frac{1}{2}\int_{0}^1 R_i(x)\frac{d^2}{dx^2}\left( x(1-x) R_j(x) \right) \frac{dx}{\sqrt{x(1-x)}},
\end{equation}
and then, the three matrices needed to reconstruct the migration term
\begin{eqnarray}
G_{ij}=\int_{0}^1 R_i(x) x R_j(x) \frac{dx}{\sqrt{x(1-x)}}, \label{mat1}\\
H_{ij}=\int_{0}^1 R_i(x) \frac{dR_j(x)}{dx}  \frac{dx}{\sqrt{x(1-x)}}, \label{mat2}\\
J_{ij}=\int_{0}^1 R_i(x) x \frac{dR_j(x)}{dx} \frac{dx}{\sqrt{x(1-x)}}. \label{mat3}
\end{eqnarray}

The matrix elements in Eqs. \eqref{RD}, \eqref{mat1}, \eqref{mat2} and \eqref{mat3} 
can be quickly determined by means of the Gauss-Chebyshev quadrature defined
in Eq. \eqref{FT}. Due to the properties of the Chebyshev polynomials
many matrix elements vanish. More particularly,
$D_{ij}$ and $J_{ij}$ are upper triangular matrices (i.e., $D_{ij}=J_{ij}=0$ if $i>j$), $H_{ij}=0$
if $i\ge j$, and $G_{ij}=0$ if $i>j+1$ or $i<j-1$. Thus, the total number of non-trivial matrix elements
that we need to compute, for a given $\Lambda$, is just $\frac{3}{2}\Lambda^2+\frac{7}{2}\Lambda-2$. This 
is much smaller than the default number of matrix elements (i.e., $\Lambda^{2K}$).

Finally, the $\Lambda^K\times\Lambda^K$ matrix elements of the corresponding $\omega$ matrix in Eq. \eqref{ode2}
can be easily recovered from the tensor product structure of the $\Lambda^K$-dimensional vector space that defines
the Chebyshev expansion (as in Eq. \eqref{expa1}). Thus, $\omega_{i_1\ldots i_P,j_1\ldots j_K}=$
\begin{eqnarray}
\label{coeffs_FP}
\sum_a \frac{1}{2N_{e,a}}D_{i_a,j_a} - \sum_{a,b}m_{ab}\left(
H_{i_a,j_a}G_{i_b,j_b}-\delta_{i_a,j_a}\delta_{i_b,j_b}-J_{i_a,j_a}\delta_{i_b,j_b}
\right),
\end{eqnarray}
with $\delta_{ij}=1$, if $i=j$, and $\delta_{ij}=0$ if $i\neq j$.

\subsection{Influx of Mutations}

The inhomogeneous terms in Eq. \eqref{migration_selection2} that model the influx of mutations are given
by effective mutation densities. As we show in the appendices, a model of mutations given by an
exponential distribution will give the same results, up to an exponentially small deviation, as a standard model 
with a Dirac delta. The motivation for using smooth effective mutation densities is that they are better 
approximated by truncated Chebyshev expansions. As we briefly explained in the review on Chebyshev polynomials 
and its truncated expansions, the convergence of a truncated expansion depends strongly on the analytical properties of the function
to be approximated. As Dirac deltas are not smooth functions, their truncated Chebyshev expansions
present bad convergence properties. This is related to the problem of Gibbs phenomena, and we will give a more detailed
account of its origin below (see \emph{Sources of error and limits of numerical methods}).

\begin{figure}[hpbt]
\begin{center}
\vspace{0.5cm}
\begin{tabular}{cc}
    \includegraphics[width=6.1cm]{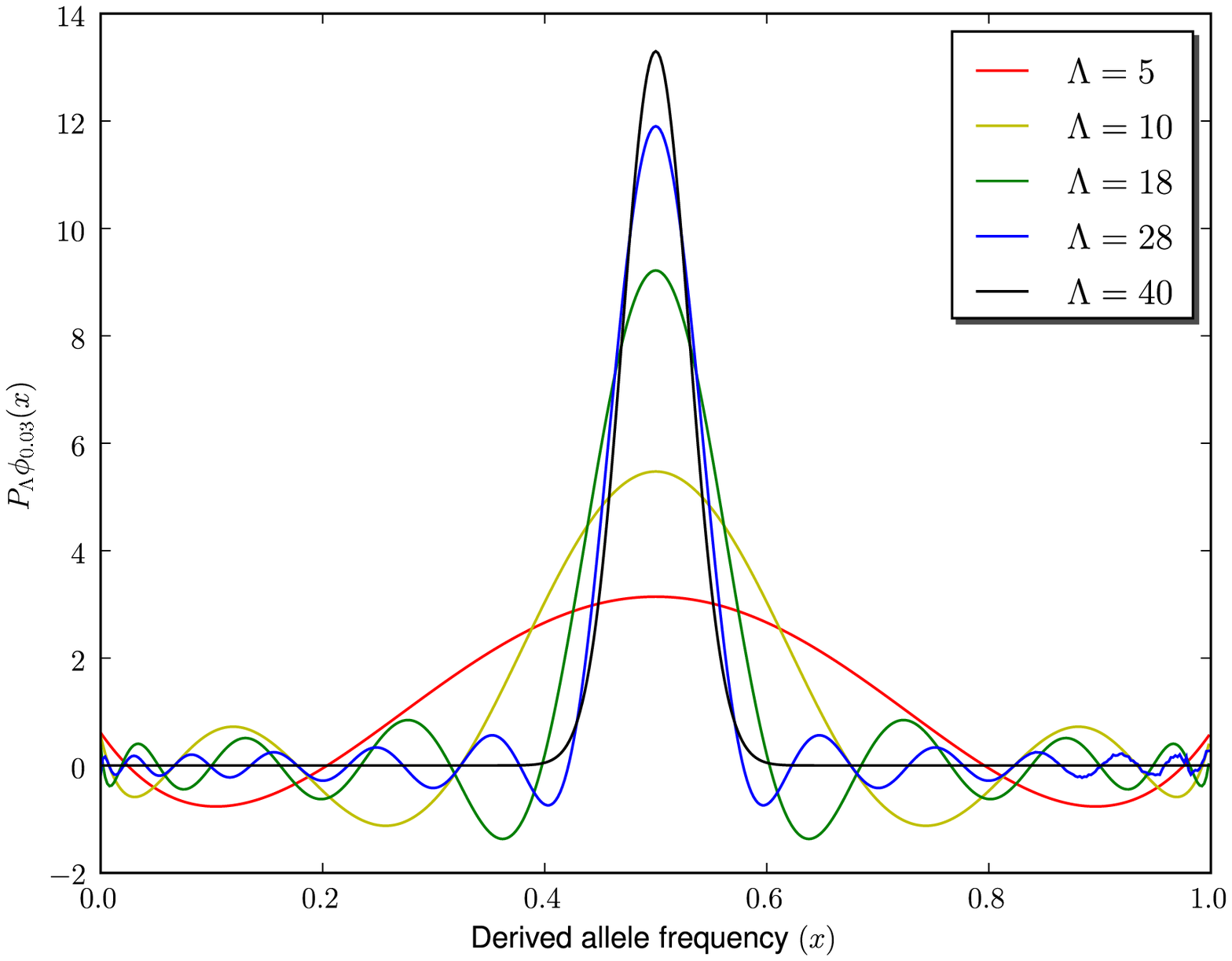}&
    \includegraphics[width=6.1cm]{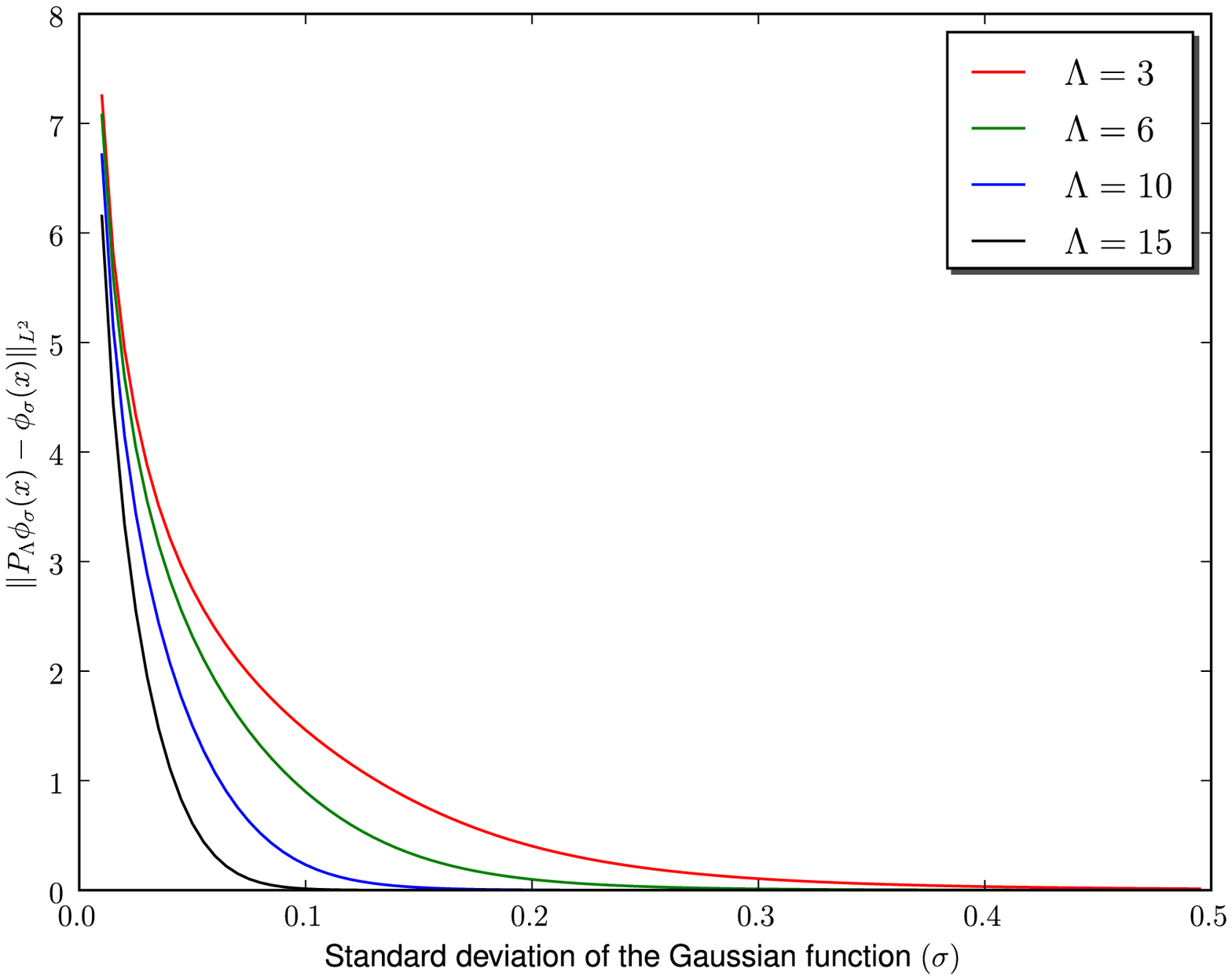}
\end{tabular}
\caption{\label{figsobolev}
On the left, we show the plot of five different truncated Chebyshev expansions for a Gaussian peaked at $x=0.5$ and $\sigma=0.03$.
On the right, we show the truncation error of different Chebyshev expansions (with $\Lambda=3,\,6,\,10$ and $15$) of a family of 
Gaussian functions peaked at $x=0.5$ and parametrized by the standard deviation $0.01\leq\sigma\leq 0.5$.
}
\end{center}
\end{figure}
In this paper, we only consider a positive influx of mutations in boundary components of dimension one. 
In order to approximate the behavior under a mutation term given by a Dirac delta, an effective mutation 
density $\mu(x)$ has to satisfy the following:
\begin{itemize}
\item The truncation error is bounded below
the established parameter, $\epsilon$, for the control of error; i.e. , $\Vert \mu(x) - \mathcal{P}_{\Lambda}\mu(x)\Vert_{L^2} < \epsilon$.
\item The expected frequency of the mass mutation-function is $\mathbb{E}_{\mu}(x)=\frac{1}{2N_e}$.
\item The mutation-function is nearly zero for relatively large frequencies (e.g., $x>0.05$), and it is as peaked as possible near $x=1/(2N_e)$.  
\end{itemize}
While the first and third qualitative requirements are straightforward, the second numerical condition is not. 
One can interpret this requirement as equivalent to fixing the neutral fixation rate to be $u$, 
because the probability that an allele at frequency $x=p$ reaches fixation at $x=1$ is $p$. 
Thus, the average number of new mutants that reach fixation per generation is 
$2N_e u \times \mathbb{E}_{\mu}(x)=u$.
This constraint can also be derived by studying the properties of the equilibrium density associated 
with this stochastic process. At equilibrium, the density $\phi_{e}(x)$ of 
derived alleles obeys
\begin{equation}
\frac{1}{4N_e}\frac{d^2 \psi}{dx^2}=-2N_e u \mu(x),
\end{equation}
with $\psi(x)=x(1-x)\phi_{e}(x)$. Therefore, the expected frequency of the mass mutation-function can be computed as
\begin{equation}
\label{first_momentum}
\int_{0}^1 x \mu(x) dx=-\frac{1}{8N_e^2 u}\int_{0}^1 x\frac{d^2 \psi}{dx^2} dx.
\end{equation}
Using the identity $x\frac{d^2 \psi}{dx^2}=\frac{d}{dx}\left[ x\frac{d\psi}{dx}\right] -\frac{d\psi}{dx}$ one can rewrite
Eq. \eqref{first_momentum} as
\begin{equation}
\int_{0}^1 x \mu(x) dx=-\frac{1}{8N_e^2 u}\frac{d \psi}{dx}(x=1).
\end{equation}
On the other hand, the probability flux associated with the equilibrium density of alleles at the boundary 
$x=1$ is $j(1)=-\frac{1}{4N_e}\psi^{\prime}(1)$. We use this to write the expected frequency of the mutation density as:
\begin{equation}
\label{flux_momentum}
\int_{0}^1 x \mu(x) dx=\frac{1}{2N_e u}j(1).
\end{equation}
In neutral evolution the probability flux at the boundary $x=1$ equals the fixation rate, which satisfies $j(1)=u$. 
Therefore, Eq. \eqref{flux_momentum} has to satisfy
\begin{equation}
\label{flux_momentum2}
\int_{0}^1 x \mu(x) dx=\frac{1}{2N_e u}j(1)= \frac{1}{2N_e u}u=\frac{1}{2N_e},
\end{equation}
which is what we wanted to show.

Numerical experiments show that for a large class of functions $\mu(x)$, and in the frequency range $x>x_\ast$, 
the associated solutions to the different diffusion problems are identical (up to a very small deviation) to the 
standard model with a Dirac delta.
$x_\ast$ is a very small frequency that depends on the choice of 
$\mu(x)$, and generally can be made arbitrarily small. It is in the region of the frequency space with  
$0\leq x \leq x_\ast$, where the behavior of the different diffusion problems can deviate most. 

The truncation error in the Chebyshev expansion depends on the smoothness of the function, and the choice
of truncation parameter (see Fig. \ref{figsobolev} for an example). For the effective mutation density $\mu(x)$, 
we use the exponential function 
\begin{eqnarray}
\mu(x)=\frac{1}{2N_e}\times \frac{\kappa^2}{1-\exp(-\kappa)-\kappa\exp(-\kappa)}\exp(-\kappa x) \label{muta_exp},
\end{eqnarray}
where the values for $\kappa(\Lambda,\epsilon)\gg 1$ are determined by saturating the bound on error:
$\Vert \mu(x) - \mathcal{P}_{\Lambda}\mu(x)\Vert_{L^2} < \epsilon$.

\subsubsection{Comparison of different mutation models at equilibrium}

We derive in the Appendix A the associated equilibrium distributions of derived alleles. 
For a model with a mutation density given by a Dirac delta, one finds the equilibrium density
\begin{equation}
\label{eq_dirac_delta}
\phi_e(x)=\frac{4N_e(2N_e-1)u x-8N_e^2 u(x-1/(2N_e))\theta(x-1/(2N_e))}{x(1-x)},
\end{equation}
with $\theta(y)$ the Heaviside step function ($\theta(y)=0$ for $y<0$, $\theta(y)=1/2$ for $y=0$, and 
 $\theta(y)=1$ for $y>0$). Which in the region $x>1/(2N_e)$ simplifies to 
\begin{equation}
\label{eq_dirac_delta2}
\phi_e(x)=\frac{4N_e u}{x}.
\end{equation}
In the case of $\mu_2(x)=c\exp(-\kappa x)$, the corresponding equilibrium density is
\begin{equation}
\label{eq_exp}
\phi_e(x)=\frac{4N_e u (1-x) +\frac{8N_e^2 u c}{\kappa}\left(
 \exp(-\kappa)(1+1/\kappa) -\exp(-\kappa) x -\frac{1}{\kappa}\exp(-\kappa x)
\right)
}{x(1-x)}.
\end{equation}

\begin{figure}[hpbt]
\begin{center}
\vspace{0.5cm}
\begin{tabular}{ccc}
    \includegraphics[width=3.6cm]{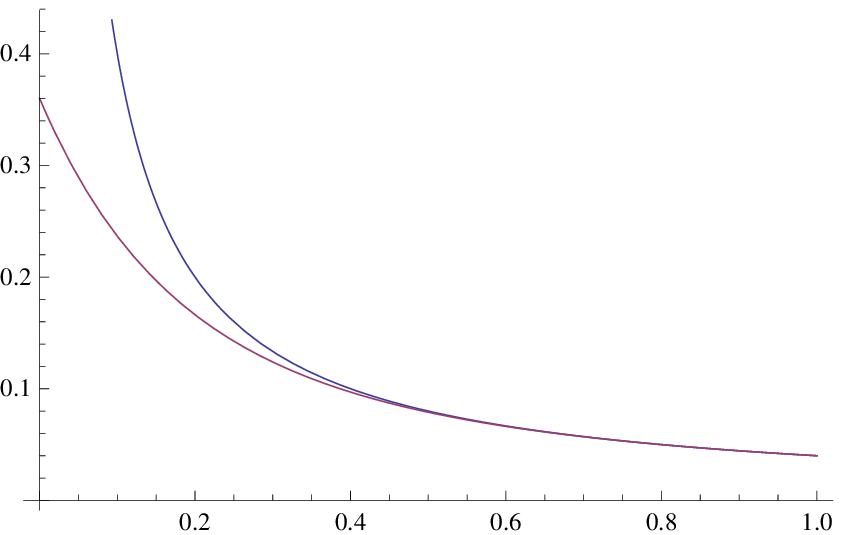}&
    \includegraphics[width=3.6cm]{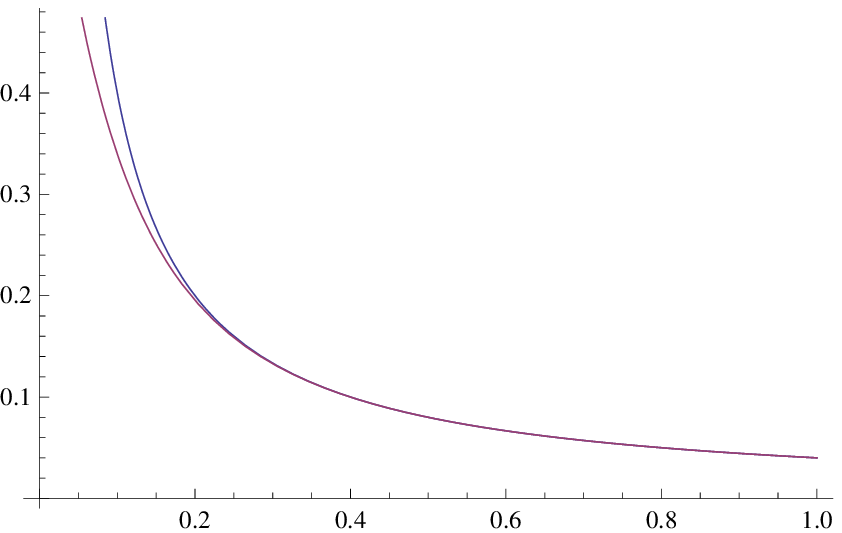}& 
    \includegraphics[width=3.6cm]{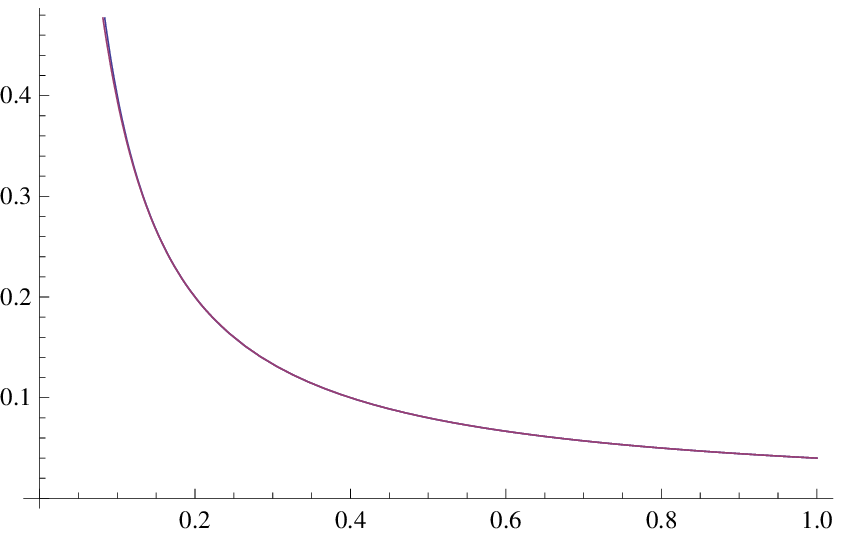} 
\end{tabular}
\caption{\label{mutation_densities_convergence}
Three comparisons of the equilibrium densities associated with 
the exponential mutation density (blue) for several values of $\kappa$ vs.
the equilibrium density associated with the Dirac delta mutational model (red). For illustrative purposes, the 
population size used was $N=10,000$ and the spontaneous mutation rate is $u=10^{-6}$.
On the left the equilibrium density associated with the exponential distribution with $\kappa=10$ is shown, in the middle 
$\kappa=20$, and on the right $\kappa=40$. For a truncation parameter $\Lambda=20$, one can choose mutation densities with 
$\kappa$ up to $43$, while keeping the truncation error below sensible limits.
}
\end{center}
\end{figure}

Therefore, a pairwise comparison of both equilibrium densities shows that the deviation from
both models when $x>x_\ast=\kappa^{-1}$ is exponentially small when equilibrium is reached 
(see Fig. \ref{mutation_densities_convergence}). 
We can show that the same is true in non-equilibrium.

\subsubsection{Non-equilibrium dynamics with effective mutation densities}

Here, we show how the non-equilibrium dynamics of a diffusion system under an exponential distribution mutation 
influx is the same (up to an exponentially small deviation) as a system where mutations enter the population 
through the standard Dirac delta $\delta(x-1/(2N_e))$, as long as the allele frequencies are bigger than certain
minimum frequency $x_\ast$. Below $x_\ast$ the dynamics will be sensitive to differences in the 
mutation densities.

Let $\varphi(x)$ be an arbitrary initial density of alleles. Let $\phi_1(x,t)$ be the solution to the diffusion 
equations under pure random drift and a mutation influx given by $\delta(x-1/(2N_e))$. $\phi_2(x,t)$ is 
the solution of the diffusion equations under pure random drift and mutation influx given by the exponential 
effective mutation density Eq. \eqref{muta_exp}. In Appendix B, we prove the following identity
in the large $t$ limit
\begin{equation}
\label{error_time}
\int_{0}^1 \vert  \phi_1(x,t) -\phi_2(x,t)\vert x(1-x)dx = \frac{4 N_e u}{\kappa}(1-\exp(-t/(2N_e)))+O(\exp(-\kappa)),
\end{equation}
with $\phi_1(x,0)=\phi_2(x,0)=\varphi$, and $\kappa\leq 2N_e$.
If we normalize
Eq. \eqref{error_time} by
$$
\lim_{t\to\infty} \int_{0}^1 \vert \phi_1(x,t) \vert x(1-x) dx,
$$
the normalized deviation of $\phi_2(x,t)$ from $\phi_1(x,t)$ is, for large $t$,
\begin{equation}
\frac{\int_{0}^1 \vert  \phi_1(x,t) -\phi_2(x,t)\vert x(1-x)dx}{\lim_{t\to\infty}\int_{0}^1 
\vert\phi_1(x,t)\vert x(1-x)dx} = \frac{2}{\kappa}(1-\exp(-t/(2N_e)))+O(e^{-\kappa},N_e^{-1}).
\end{equation}
We can also show, by applying the Minkowski inequality to Eq. \eqref{deviation_evolution} in Appendix B, that
\begin{equation}
\frac{\int_{0}^1 \vert  \phi_1(x,t) -\phi_2(x,t)\vert x(1-x)dx}{\lim_{t\to\infty}\int_{0}^1 
\vert\phi_1(x,t)\vert x(1-x)dx} \leq \frac{2}{\kappa}(1+\exp(-t/(2N_e)))+O(e^{-\kappa},N_e^{-1}),
\end{equation}
for all $t>0$. This means that the deviation is bounded by $O(\kappa^{-1})$ for all $t$, and therefore
the non-equilibrium dynamics of $\phi_1(x,t)$ and $\phi_2(x,t)$ are identical in the large $\kappa$ limit.

As $\vert \phi_1(x,0)-\phi_2(x,0)\vert=0$ at time zero and
the deviation of $\phi_2(x,t)$ from $\phi_1(x,t)$ attains equilibrium in the large $t$ limit, we can study 
the frequency dependence of such deviation by looking at the equilibrium
\begin{equation}
\lim_{t\to\infty} \phi_1(x,t) -\phi_2(x,t) = \frac{4N_e u e^{-\kappa x}}{x(1-x)} + O(e^{-\kappa}).
\end{equation}
Here, the $O(e^{-\kappa})$ term exactly cancels the singularity at $x=1$ and the deviation decays exponentially as
a function of the frequency. This shows that for frequencies $x>x_\ast = \kappa^{-1}\geq 1/(2N_e)$ the 
dynamics of a model with mutation influx given by a Dirac delta is the same, up to an exponentially small deviation,
as the non-equilibrium dynamics of a model with exponential mutation density.

\subsection{Branching-off of populations}

Modeling a population splitting event also involves the use of Dirac deltas, as in Eq. \eqref{founderII}, or peaked functions
such as Eq. \eqref{founder}, whose truncated Chebyshev expansions may present bad convergence properties. These Gibbs-like phenomena can be dealt
with in a similar way as we did with the mutation term of Eq. \eqref{migration_selection2}. 

We implemented two different solutions to this problem and both solutions yielded similar results.
\begin{figure}[hpbt]
\begin{center}
\vspace{0.5cm}
    \includegraphics[width=8.5cm]{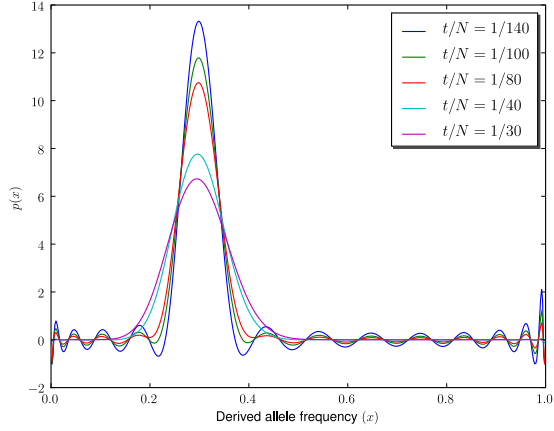}
\caption{\label{figii}
Diffusion under pure random drift acts by smoothing out the initial density at $t=0$. Here we show  
numerical solutions to the diffusion equations with $\Lambda=28$, at $5$ different times, with initial condition
$\phi(x,t=0)=\delta(x-0.3)$. As time passes, the numerical solution approaches the exact solution more quickly, and the Gibbs phenomena
disappear.
}
\end{center}
\end{figure}
First, we constructed a smoothed approximation of the Dirac delta by using Gaussian functions:
\begin{equation}
\label{thick_delta}
\tilde{\delta}(x_a-x_{K+1})=\frac{1}{w(x_a)}\exp\left( -\frac{(x_a-x_{K+1})^2}{2 \sigma(x_a)^2} \right),
\end{equation}
with 
\begin{equation}
\label{total_mass}
w(x_a)=\int_0^1 \exp\left( -\frac{(x_a-x_{K+1})^2}{2 \sigma(x_a)^2} \right) dx_{K+1},
\end{equation}
and $\sigma(x_a)$ as a standard deviation which is chosen as small as possible while preserving the bound on error,
$\Vert \tilde{\delta}(x_a-x) - \mathcal{P}_{\Lambda}\tilde{\delta}(x_a-x)\Vert_{L^2[0\leq x\leq 1]} < \epsilon$, 
for any value of $ x_a\in [0,1]$. In order to map the $\tilde{\delta}$-function in Eq. \eqref{thick_delta} to a truncated Chebyshev expansion, 
\begin{equation}
\label{gauss_chebyI}
\mathcal{P}_{\Lambda}\tilde{\delta}(x_a-x_{K+1})=\sum_{i=0}^{\Lambda-1}\sum_{j=0}^{\Lambda-1}\Delta_{ij}
R_i(x_a)R_j(x_{K+1}),
\end{equation}
one has to perform a Gauss-Chebyshev quadrature in 2 dimensions, $0\leq x_a \leq 1$, $0\leq x_{K+1} \leq 1$:
\begin{equation}
\label{gauss_chebyII}
\Delta_{ij}= \int_{0}^1\int_{0}^1 \tilde{\delta}(x_a-x) R_i(x_a)R_j(x)\frac{dx_a}{\sqrt{x_a(1-x_a)}}
\frac{dx}{\sqrt{x(1-x)}}.
\end{equation}
  
The second approach exploits the analytical behavior of diffusion under pure random drift (i.e., with no migration). By
Kimura's solution to the diffusion PDE in terms of the Gegenbauer polynomials $\{ \mathcal{G}_i(z) \} $, see \cite{kimuraI}, 
we know that the time evolution of 1-d density is
\begin{equation}
\label{kimuraeq}
\mathcal{P}_{\Lambda}\phi(x,t)=\sum_{i=0}^{\Lambda-1} \frac{2(i+1)+1}{(i+1)(i+2)}(1-r^2)\mathcal{G}_i(r)\mathcal{G}_i(z)\exp(-(i+1)(i+2)t/4N ),
\end{equation}
with $r=1-2p$, $z=1-2x$ and $\phi(x,0)=\delta(x-p)$. Thus, in the exact solution to the diffusion
equation, the time evolution of the coefficients of degree $i$ in the Gegenbauer expansion is described by the term 
$\exp(-(i+1)(i+2)t/4N)$. This means that diffusion smooths out
the Dirac delta at initial time. Fig. \ref{figii} represents the evolution of the density at different times. 

Thus, we can use diffusion under pure random drift to smooth out the density introduced after the population splitting event. 
Let $\phi_{K}(x,t)$ be the density before the splitting and let $a$ be 
the population from which population $K+1$ is founded. We initially consider the density
\begin{equation}
\phi_{K+1}(x, x_{K+1},t^\prime =0) = \phi_{K}(x,t) \delta(x_a-x_{K+1}).
\end{equation}
The associated Chebyshev expansion $\mathcal{P}_{\Lambda}\phi_{K+1}(x, x_{K+1},t)$ will present Gibbs-phenomena. However, by 
diffusing for a short time $\tau$ under pure random drift
\begin{equation}
\frac{\partial}{\partial t} \phi_{K+1}(x, x_{K+1}, t)= \sum_{b=1}^{K+1}
\frac{1}{2}\frac{\partial^2}{\partial x_{b}^2}\left( \frac{x_b(1-x_b)}{2S_b} \phi_{K+1}(x, x_{K+1},t)\right),
\end{equation}
(with $S_a=S_{K+1}=W$, $S_b=V$ for $K+1\neq b\neq a$, and $V\gg W$), $\phi_{K+1}(x, x_{K+1},\tau)$ becomes tractable
under Chebyshev expansions. In other words, by choosing $\tau$ such that the error bound is satisfied
$$
\Vert  \phi_{K+1}(x, x_{K+1},\tau) - \mathcal{P}_{\Lambda}\phi_{K+1}(x, x_{K+1},\tau) \Vert_{L^2} < \epsilon,
$$
we obtain a smooth density after the population splitting event which can follow the regular diffusion with migration
prescribed in the problem, and approximate accurately the branching-off event. In some limits this approximation can fail,
though we leave the corresponding analysis for the next section.

Here, we do not consider the numerical solution to the problem of splitting with admixture, although we are confident that
it should be possible to solve along similar arguments.

\subsection{Sources of error and limits of numerical methods}

There are two major sources of error in these numerical methods. First, the solution of the diffusion equation is itself 
a time-continuous approximation to the time evolution of a probability density evolving under a discrete Markov chain. Hence,
whenever the diffusion approximation fails, its numerical implementation will also fail. Secondly, a numerical solution by
means of spectral expansions involves an approximation of the infinite dimensional space of functions on a domain by a 
finite dimensional space generated by bases of orthonormal functions under certain $L^2$-product. As we show below, 
under a broad set of conditions the numerical solution will converge accurately to the exact solution; otherwise, the 
numerical solution can fail to approximate the exact solution. A third source of error appears because one has to discretize 
time in order to integrate the high-dimensional ODE that approximates the PDE. 
Fortunately, this source of error can be ignored because the diffusion generators yield a stable time evolution.

We summarize below the main conditions that have to be satisfied in order to obtain high-quality numerical solutions
to the PDEs studied in this work.

\subsubsection{Limits of diffusion equations}
In the diffusion approximation to a Markov chain, the transition probability is approximated by a Gaussian
distribution \cite{ewens}. Here, we review the derivation of the diffusion equation as the continuous limit 
of a Markov chain, in order to emphasize the assumptions made and determine the limits of this approximation.

Given a Markov process defined by a discrete state space $\mathcal{S}$, 
transition matrices $p(I\vert J)$, initial value $K\in\mathcal{S}$ and discrete time $\tau=0,1,\ldots $, 
the probability that the state will be at $I$ at time $\tau$ is $f(I\vert K,\, \tau)$, where $f(I\vert K,\, \tau)$ 
obeys the recurrence relation
\begin{equation}
\label{markov}
f(I\vert K,\, \tau)=\sum_{J\in \mathcal{S}} p(I\vert J) f(J\vert K,\, \tau-1).
\end{equation}
In the diffusion approximation one considers a sequence of discrete state spaces 
$\{ \mathcal{S}_\lambda \}_{\lambda\in \mathbb{Z}_+}$ such that in the 
limit $\lambda\to\infty$ the state space $\mathcal{S}_\infty$ converges to a smooth manifold (in practical applications,
a compact domain $D\subset\mathbb{R}^K$). 

In this paper, we take $\mathcal{S}_\lambda$ to be $[0,\,\lambda]^K$, and $\mathcal{S}_\infty \sim [0,1]^K$. 
Therefore, the state variables can
be re-scaled as $K_a /\lambda=x_a$, with $a=1,\ldots, K$ and $K_a \in [0,\,\lambda]^K$. Similarly, we introduce the time
variable $t=\tau/\lambda$. In the large $\lambda$ limit, the transition probability for the change of the state
from time $\tau/\lambda$ to time $(\tau+1)/\lambda$ is governed by a distribution with moments
\begin{eqnarray}
\mathbb{E}(\delta x_a\vert x)=M_a(x)/\lambda + O(1/\lambda^2), \label{mean}\\
\mathbb{E}(\delta x_a\delta x_b\vert x) - \mathbb{E}(\delta x_a\vert x)\mathbb{E}(\delta x_b\vert x)=C_{ab}(x)/\lambda + O(1/\lambda^2), \\
\mathbb{E}(\delta x_a^3\vert x)=O(1/\lambda^2).
\end{eqnarray}
In this continuous limit, the equation that describes the time evolution of the Markov chain in Eq. \eqref{markov} can be written as
a forward Kolmogorov equation if we neglect terms of order $O(1/\lambda^2)$. However, if $M_a(x)$ is proportional to 
$\lambda$, the $O(1/\lambda^2)$ terms in Eq. \eqref{mean} cannot be neglected and the diffusion approximation will not be valid. As in 
this paper we take $\lambda = 2N_e$, and $M_a(x)$ proportional to the migration rates $m_{ab}$, if the
migration rates obey $2N_{e,a}m_{ab}\leq O(1)$ the diffusion approximation will be valid.
Indeed, computer experiments show that the numerical solutions become unstable and yield incorrect results if this bound is violated.
This limit precisely defines when two populations can be considered the same \cite{heyI}. Therefore, in cases
when $2N_{e,a}m_{ab}\gg O(1)$, we can consider populations $a$ and $b$ as two parts of the same population.
Another assumption in the diffusion approximation is that a binomial distribution with $2N_{e,a}$ degrees of freedom 
can be approximated by a Gaussian distribution. This will be a valid approximation as long as $N_{e,a}$ is large enough.
Numerical experiments show that the approximation is accurate if $N_{e,a}>100$; otherwise, effects due to the 
finiteness of the Markov chain cannot be neglected and the approximation will fail.
\begin{figure}[hpbt]
\begin{center}
\vspace{0.1cm}
    \includegraphics[width=2.4cm]{my_image2.epsi}
\end{center}
\end{figure}

\subsubsection{Limits of spectral expansions}

Spectral methods, as with any numerical scheme for solving PDEs, require several assumptions about the behavior of the solution
of the PDE. The most important one is that one can approximate the solution as a series of smooth basis functions,
\begin{equation}\label{expansion}
\mathcal{P}_\Lambda \phi(x,t)=\sum_{i_1 =0}^{\Lambda-1}\cdots\sum_{i_P=0}^{\Lambda-1} 
\alpha_{i_1,i_2,\ldots i_P}(t)T_{i_1}(x_{1})T_{i_2}(x_{2})\cdots T_{i_P}(x_{P}).
\end{equation}
In other words, the projection of the solution $\mathcal{P}_\Lambda \phi(x,t)$ is assumed to approximate $\phi(x,t)$ well in some
appropriate norm for sufficiently large $\Lambda$. As one has to choose finite values for $\Lambda$, Eq.\eqref{expansion} will sometimes
fail to approximate correctly the solution of the PDE. 

In the applications of this paper, the basis of functions that we use consist of the Chebyshev polynomials of the first kind\footnote{
One can work either 
with the basis of functions $\{ T_i(x) \}$ on $x\in[-1,1]$, or with the 
re-scaled basis $\{ R_i(x) \}$ defined on $x\in[0,1]$, by performing a simple scale transformation.}. Below we provide bound estimates 
for the truncation error $\Vert \mathcal{P}_\Lambda \phi(x,t) - \phi(x,t) \Vert_{L^2[-1,1]^K} $,
to understand the quality of the approximate solutions for different values of $\Lambda$, (see also \cite{spectral, canuto} 
for different choices of basis functions).

More precisely, as the $L^2$ inner product and norm in the Chebyshev problem are:
\begin{equation}\label{l2inner}
\langle f,\, g \rangle_{L^2[-1,1]^K}=\int_{[-1,1]^K}f(x)g(x)\prod_{i=1}^K\frac{dx_i}{\sqrt{1-x_i^2}},
\end{equation}
and
\begin{equation}\label{l2norm}
\Vert f\Vert_{L^2[-1,1]^K}^2=\int_{[-1,1]^K}\vert f(x)\vert^2 \prod_{i=1}^K\frac{dx_i}{\sqrt{1-x_i^2}},
\end{equation}
the terms in the expansion Eq. \eqref{expansion} can be computed by performing inner products
\begin{equation}\label{coeffs}
\alpha_{i_1,i_2,\ldots i_K}(t)=\left( \frac{2}{\pi} \right)^K \int_{[-1,1]^K}\phi(x,t) 
T_{i_1}(x_{1})T_{i_2}(x_{2})\cdots T_{i_K}(x_{K}) \prod_{j=1}^K\frac{dx_j}{c_{i_j}\sqrt{1-x_j^2}},
\end{equation}
with $c_0=2$ and $c_j=1$ ($j>0$). A consequence of the orthogonality of the basis functions is that
the squared truncation error admits a simple formulation in terms of the coefficients in the expansion:
\begin{equation}\label{trunc}
\Vert \mathcal{P}_\Lambda \phi(x,t) - \phi(x,t) \Vert^2_{L^2[-1,1]^K}= 
\left( \frac{2}{\pi} \right)^K \sum_{i_1\geq\Lambda}\cdots\sum_{i_K\geq\Lambda}   \vert \alpha_{i_1,i_2,\ldots i_K}(t) \vert^2.
\end{equation}
Thus, the truncation error depends only on the decay of the higher modes $\vert \alpha_{i_1,i_2,\ldots i_K} \vert$ in the expansion.
On the other hand, the decay of these higher modes depends on the analytical properties of $\phi(x,t)$ itself.
For instance, if $\phi(x,t)\in C^{2q_1-1,2q_2-1,\ldots,2q_K-1}\left( [-1,1]^K \right)$, i.e. if 
\begin{equation}\label{bound1} 
\left\Vert  \left( \frac{\partial}{\partial x_1}\right)^{2q_1-1}\left( \frac{\partial}{\partial x_2}\right)^{2q_2-1}\cdots 
\left(\frac{\partial}{\partial x_K}\right)^{2q_K-1} \phi(x,t) \right\Vert_{L^2[-1,1]^K} <\infty, 
\end{equation}
we can integrate by parts Eq. \eqref{coeffs}, as we did in Eq. \eqref{conve2}, to write the decay of each mode as
\begin{eqnarray}
\label{decay1}\\
\vert \alpha_{i_1,i_2,\ldots i_K}(t) \vert=\left( \frac{2}{\pi} \right)^K 
\Bigg\vert \int_{[-1,1]^K} \left[ \prod_{j=1}^K \left(
\sqrt{1-x_j^2}\frac{\partial}{\partial x_j}\right)^{2q_j} 
\phi(x,t)\right]\times \nonumber \\ \frac{T_{i_1}(x_{1})}{c_{i_1} i_1^{2q_1} \sqrt{1-x_1^2}} dx_1\cdots 
\frac{T_{i_K}(x_{K})}{c_{i_K} i_K^{2q_K}\sqrt{1-x_K^2}} dx_K
\Bigg\vert.\nonumber
\end{eqnarray}
Eq. \eqref{decay1} implies that the truncation error is directly related to the smoothness of $\phi(x,t)$; it follows
that we can bound the truncation error as a function of $\Lambda$:
\begin{equation}\label{bound2} 
\Vert \mathcal{P}_\Lambda \phi(x,t) - \phi(x,t) \Vert_{L^2[-1,1]^K} \leq  C(q) \Lambda^{-\sum_j q_j} \left\Vert 
\prod_{j=1}^K \left[\sqrt{1-x_j^2}\frac{\partial}{\partial x_j}\right]^{q_j} \phi(x,t)
\right\Vert_{L^2[-1,1]^K}.
\end{equation}
Another convenient measure of smoothness is the Sobolev norm:
\begin{equation}\label{sobolev} 
\left\Vert \Phi(x) \right\Vert_{W^{q_1,\ldots, q_K}[-1,1]^K}^2 = \sum_{s_1=0}^{q_1}\cdots \sum_{s_P=0}^{q_K}
 \left\Vert  
\prod_{j=1}^K \left(\frac{\partial}{\partial x_j}\right)^{s_j}
\Phi(x) \right\Vert_{L^2[-1,1]^K};
\end{equation}
in terms of the Sobolev norm, the truncation error is bounded as
\begin{equation}\label{bound3} 
\Vert \mathcal{P}_\Lambda \phi(x,t) - \phi(x,t) \Vert_{L^2[-1,1]^K} \leq  C \Lambda^{-\sum_j q_j} \left\Vert 
\phi(x,t)
\right\Vert_{W^{q_1,\ldots, q_K}[-1,1]^K}.
\end{equation}
A corollary of Eq. \eqref{bound3} is that if $\phi(x,t)$ is smooth, $\mathcal{P}_\Lambda \phi(x,t)$ converges to
$\phi(x,t)$ more rapidly than any finite power of $\Lambda^{-1}$. This is indeed the basic property that has given
name to spectral methods.

In the absence of influx of polymorphisms in the populations, the time evolution of the density obeys pure 
diffusion, and therefore $\vert \alpha_{i_1,i_2,\ldots i_K}(t) \vert \to 0$ when
$t\to\infty$ as it follows from Eq. \eqref{kimuraeq}. This means that diffusion acts as a 
smoothing operator on the initial density.
Empirically, we find that in the presence of influx of polymorphisms the density can also be approximated
by spectral expansions and the truncation error remains low.

After two populations split and the $K$-dimensional diffusion becomes a $K+1$ dimensional process,
the $K+1$ dimensional density becomes a distribution concentrated in the linear subspace of $[-1,1]^{K+1}$ defined by
$x_a=x_{a+1}$ (with $a$ and $a+1$ labeling the two daughter populations that just split). Such density has an infinite
Sobolev norm and cannot be represented as a finite sum of polynomials. Fortunately, the diffusion generator acts on 
the density by smoothing it out and by bringing the density to a density with finite Sobolev norm. The main variables involved
in this process are: the time difference between the current splitting event and the next one, $T_{A+1}-T_A$ , and the
effective population sizes $N_{e,a}$ and $N_{e,a+1}$ of the daughter populations. Therefore, depending on the choice of
the truncation parameter $\Lambda$, a minimum diffusion time $t_m(N_{e,a}, N_{e,a+1}, \Lambda)$ will be necessary
to bring the truncation error within desired limits $\Vert \mathcal{P}_\Lambda \phi(x,t_m) - \phi(x,t_m) \Vert_{L^2}\leq\epsilon$. Here,
$\epsilon$ is the control parameter on numerical error. Therefore, the bigger the largest effective population size of the two daughter
populations, the bigger will be such minimum diffusion time. If the time difference between the current splitting event and the next one
is bigger than 
\begin{equation}\label{min}
t_m(N_{e,a}, N_{e,a+1}, \Lambda)=C(\Lambda)\, \mathrm{max}(N_{e,a}, N_{e,a+1}),
\end{equation}
(where $C(\Lambda)$ is a function that can be computed numerically),
the resulting numerical error will stay below the desired limits. 
\begin{figure}[hpbt]
\begin{center}
\vspace{0.1cm}
    \includegraphics[width=6.128cm]{my_image3.epsi}
\end{center}
\end{figure}
As our model aims to reproduce the real SNP Allele Frequency Spectrum density 
there should exist low error approximations of such density (that we denote as $\hat{\gamma}(x)$)
in terms of polynomial expansions. 
Otherwise, the methods here presented will fail to solve the problem. This can only 
happen if $\hat{\gamma}(x)$ is so rugged, i.e. the corresponding Sobolev norm is so high, that the largest finite choice
for $\Lambda$ that we can implement in our computer-code is not large enough to approximate accurately $\hat{\gamma}(x)$:  
\begin{equation}\label{bound4} 
\Vert \mathcal{P}_{\Lambda_{max}} \hat{\gamma}(x) - \hat{\gamma}(x) \Vert_{L^2[-1,1]^K} \sim 
C \Lambda_{max}^{-\sum_j q_j} \left\Vert 
\hat{\gamma}(x)
\right\Vert_{W^{q_1,\ldots, q_K}[-1,1]^K}\gg \epsilon .
\end{equation}
In case that Eq. \eqref{bound4} is obeyed, it is likely that $2$ or more populations are so closely related that 
we can treat them as if they were one identical population. If we reduce the dimensionality of the problem in this way 
(by only incorporating differentiated populations), the correlations will disappear and the Sobolev norm of $\hat{\gamma}^\prime(x)$ will 
be such that we will be able to find a sensible parameter $\Lambda$ to approximate $\hat{\gamma}^\prime(x)$
as a truncated Chebyshev expansion.

\section{Conclusion}

In this paper we have introduced a forward diffusion model of the joint allele frequency spectra, and a numerical
method to solve the associated PDEs. Our approach is inspired by recent work in which similar
models were proposed \cite{williamson, evans, gutenkunst}. Analogously,
our methods are quite general and can accomodate selection coefficients and
time dependent effective population sizes. 

The major novelties of the model here presented with respect to previous work are:
\begin{itemize}
\item The introduction of spectral methods/finite elements in the context of
forward diffusion equations and infinite sites models. Traditionally, these techniques yield
better results than finite differences schemes when the dimension of the domain is high 
(i.e., when the final number of populations is high),
and the solutions are smooth. A comparison of our implementation using spectral methods,
and previous implementations using finite differences \cite{gutenkunst}, will be the matter of future work.
\item A set of boundary conditions that deals with the possibility that some 
polymorphisms reach fixation in some populations while remaining polymorphic in other populations. When the 
differences in effective population sizes between different populations  
are large, this phenomenon can become very important. Here, we have introduced a solution to address this
possible scenario. Previous work imposed zero flux at the boundaries \cite{gutenkunst}, and hence avoided 
the fixation of polymorphisms in some populations while remaining polymorphic in the rest.
\item The introduction of effective mutation densities, which generalize previous models for the influx of mutations
\cite{evans}. We have emphasized
how different ways to inject mutations at very low frequencies converge to the same solution for 
larger frequencies.
\end{itemize}

The non-equilibrium theory of Allele Frequency Spectra is of primary importance to analyze population
genomics data. Although it does not make use of information about haplotype structure or linkage
non-equilibrium, the analysis of AFS allows the study of demographic history and the inference of natural selection.
In this work, we have extended the diffusion theory of the multi-population AFS, to accommodate spectral methods, a general
framework for the influx of mutations, and non-trivial boundary interactions.

\section{Acknowledgements}

We thank Rasmus Nielsen, Anirvan Sengupta, Simon Gravel and Vitor Sousa for helpful conversations. This work was partially supported
by the National Institutes of Health (grant number R00HG004515-02 to K. C. and grant number R01GM078204 to J. H.).

\newpage
\appendix

\section{Comparison of mutational models at equilibrium}

In this appendix we compute the equilibrium densities associated with Wright-Fisher processes with
mutation. Two types of mutation processes are considered, both modeled by 
a mutation density. The first mutation density is a Dirac delta, while the second one is an 
exponential distribution.

As the diffusion equation that describes the time evolution of the density of alleles for diallelic SNPs is
\begin{equation}
\frac{\partial \phi(x,t)}{\partial t} = \frac{1}{4N_e}\frac{\partial^2}{\partial x^2}\left[ x(1-x)\phi(x,t) \right]+2N_e u \mu(x),
\end{equation}
the equilibrium density $\phi_{e}(x)$ satisfies $\frac{\partial \phi_e(x)}{\partial t}=0$. By using instead the function
$\psi(x)=x(1-x)\phi_e(x)$, the associated second order ordinary differential equation becomes
\begin{equation}
\label{ode_second_grade}
\frac{1}{4N_e}\frac{\partial^2 \psi(x)}{\partial x^2}+2N_e u \mu(x)=0.
\end{equation}
As Eq. \eqref{ode_second_grade} is only defined for $x>0$, we can use Laplace transforms to solve the equation. 
Let 
\begin{equation}
\tilde{\mu}(s)=\int_0^\infty \mu(x) \exp(-sx)dx,
\end{equation}
be the Laplace transform associated with the mutation density $\mu(x)$, and
\begin{equation}
\int_0^\infty \frac{\partial^2 \psi(x)}{\partial x^2} \exp(-sx)dx=s^2 \tilde{\psi}(s)-s\psi(0)-\psi^{\prime}(0),
\end{equation}
the Laplace transform associated with $\psi^{\prime\prime}(x)$, with $\psi(0)$ and $\psi^{\prime}(0)$ integration constants. 
Therefore, in the $s$ domain, $\tilde{\psi}(s)$ is
\begin{equation}
\tilde{\psi}(s)=\frac{s\psi(0)+\psi^{\prime}(0)-8N_e^2 u \tilde{\mu}(s)}{s^2},
\end{equation}
and by performing the inverse Laplace transform we obtain the solution to the equilibrium density
\begin{equation}
\label{integration_equilibrium}
\phi_e(x)=\frac{1}{x(1-x)}\frac{1}{2\pi i} \lim_{T\to \infty} \int_{\epsilon-iT}^{\epsilon+iT}  
\frac{s\psi(0)+\psi^{\prime}(0)-8N_e^2 u \tilde{\mu}(s)}{s^2} \exp(sx) ds.
\end{equation}
We fix the integration constants, $\psi(0)$ and $\psi^{\prime}(0)$, by requiring $\phi_e(x)$ to be finite 
at $x=1$, and the probability flow at $x=1$ to be equal to $u$,
\begin{equation}
j(1)=-\frac{1}{4N_e}\psi^{\prime}(0)=u.
\end{equation}
As an example, we can evaluate exactly Eq. \eqref{integration_equilibrium}, for $\mu_1(x)=\delta(x-1/(2N_e))$ and
$\mu_2(x)=c\exp(-\kappa x)$. For the Dirac delta, the Laplace transform is 
\begin{equation}
\tilde{\mu}_1(s)=\exp(-s/(2N_e)).
\end{equation}
If we compute the corresponding inverse Laplace transform in
Eq. \eqref{integration_equilibrium}, and fix the integration constants as explained above, we find
the equilibrium density
\begin{equation}
\label{eq_dirac_delta}
\phi_e(x)=\frac{4N_e(2N_e-1)u x-8N_e^2 u(x-1/(2N_e))\theta(x-1/(2N_e))}{x(1-x)},
\end{equation}
with $\theta(y)$ the Heaviside step function ($\theta(y)=0$ for $y<0$, $\theta(y)=1/2$ for $y=0$, and 
 $\theta(y)=1$ for $y>0$). If $x>1/(2N)$, Eq. \eqref{eq_dirac_delta} simplifies to 
\begin{equation}
\label{eq_dirac_delta2}
\phi_e(x)=\frac{4N_e u}{x}.
\end{equation}
In the case of $\mu_2(x)=c\exp(-\kappa x)$, the Laplace transform is
\begin{equation}
\tilde{\mu}_2(s)=\frac{c}{s+\kappa},
\end{equation}
and the corresponding equilibrium density, after integrating Eq. \eqref{integration_equilibrium}, is
\begin{equation}
\label{eq_exp}
\phi_e(x)=\frac{4N_e u (1-x) +\frac{8N^2 u c}{\kappa}\left(
 \exp(-\kappa)(1+1/\kappa) -\exp(-\kappa) x -\frac{1}{\kappa}\exp(-\kappa x)
\right)
}{x(1-x)},
\end{equation}
which in the large $\kappa$ limit, and for $x\gg 1/\kappa$, converges exponentially quickly to 
\begin{equation}
\phi_e(x)=\frac{4N_e u}{x}.
\end{equation}

In the limit $x\to 0$, $\phi_e(x)$ is finite only iff 
$c=\frac{1}{2N_e}\times \frac{\kappa^2}{1-\exp(-\kappa)-\kappa\exp(-\kappa)}$, which is the normalization choice
made in Eq. \eqref{muta_exp}, and the only one satisfying
\begin{equation}
\int_{0}^1 x \mu_2(x)=\frac{1}{2N_e}.
\end{equation}
This shows how a mutation model defined by a certain class of smooth mutation densities
reaches the same equilibrium density, up to a small deviation, as the standard model with a Dirac delta.

\newpage

\section{Comparison of mutational models at non-equilibrium}

In this appendix we compare the non-equilibrium dynamics of models
with a mutation influx described by exponential distributions, with models that consider
a standard Dirac delta.

More particularly, we prove that if $\phi_1(x,t)$ is the solution to an infinite sites model PDE,
with absorbing boundaries,
\begin{equation}
\label{equation_delta}
\frac{\partial \phi_1(x,t)}{\partial t} = 
\frac{1}{4N_e}\frac{\partial^2}{\partial x^2}\left[ x(1-x) \phi_1(x,t) \right] + 2N_e u\delta(x-1/(2N_e)),
\end{equation}
and $\phi_2(x,t)$ is the solution to the same model, but with an exponential mutation density
\begin{equation}
\label{equation_exp}
\frac{\partial \phi_2(x,t)}{\partial t} = 
\frac{1}{4N_e}\frac{\partial^2}{\partial x^2}\left[ x(1-x) \phi_2(x,t) \right]
+ u \frac{\kappa^2 \exp(-\kappa x)}{1-\exp(-\kappa)-\kappa\exp(-\kappa)},
\end{equation}
then, the \emph{deviation} of $\phi_2(x,t)$ with respect to $\phi_1(x,t)$, as a function of time and for
any initial condition $\phi_2(x,t=0)=\phi_1(x,t=0)=\varphi(x)$, is, in the large $t$ limit,
\begin{equation}
\label{deviation_a}
\int_{0}^1 \vert  \phi_1(x,t) -\phi_2(x,t)\vert x(1-x)dx = \frac{4N_e u}{\kappa}(1-\exp(-t/(2N_e)))+O(e^{-\kappa}).
\end{equation}
Here, $\vert\, \cdot\, \vert$ is the absolute value, and $O(e^{-\kappa})$ are terms that decay exponentially as a 
function of $\kappa$, which can be neglected in the large $\kappa$ limit. 

As the total number of SNPs that are polymorphic in one population depends on the population size and 
the mutation rate, it is convenient to normalize the deviation Eq. \eqref{deviation_a} by
$\lim_{t\to\infty}\int_{0}^1 \vert\phi_1(x,t)\vert x(1-x)dx=(2N_e -1)u$. In this normalization we have 
\begin{equation}
\frac{\int_{0}^1 \vert  \phi_1(x,t) -\phi_2(x,t)\vert x(1-x)dx}{\lim_{t\to\infty}\int_{0}^1 
\vert\phi_1(x,t)\vert x(1-x)dx} = \frac{2}{\kappa}(1-\exp(-t/(2N_e)))+O(e^{-\kappa},N_e^{-1}).
\end{equation}

To prove Eq. \eqref{deviation_a},
we first describe the solutions to Eq. \eqref{equation_delta} and Eq. \eqref{equation_exp}. Both equations consist
of a homogeneous term and an inhomogeneous contribution given by the mutation density. As they are linear equations,
the solution to the PDE is the sum of a homogeneous and an inhomogeneous term
\begin{equation}
\phi_{1}(x,t)=\phi_1^{h}(x,t)+\phi^e_1(x),
\end{equation}
satisfying
\begin{eqnarray}
\frac{\partial \phi^{h}_1(x,t)}{\partial t} = 
\frac{1}{4N_e}\frac{\partial^2}{\partial x^2}\left[ x(1-x) \phi^h_1(x,t) \right] +
\nonumber\\
\frac{1}{4N_e}\frac{\partial^2}{\partial x^2}\left[ x(1-x) \phi^e_1(x) \right] + 2N_e u\delta(x-1/(2N_e)).
\label{equation_delta2}
\end{eqnarray}
Hence, the only time-independent term $\phi^e_1(x)$ that solves Eq. \eqref{equation_delta2} is the equilibrium density 
Eq. \eqref{eq_dirac_delta},
and $\phi^{h}_1(x,t)$ obeys a standard diffusion equation with no mutation density, and with initial condition 
$\phi^{h}_1(x,t=0)=\varphi(x)-\phi^e_1(x)$. If $\mathcal{L} \phi^h_1(x,t)$ denotes the Fokker-Planck operator
acting on $\phi^h_1(x,t)$,
$$
\mathcal{L} \phi^h_1(x,t)=\frac{1}{4N_e}\frac{\partial^2}{\partial x^2}\left[ x(1-x) \phi^h_1(x,t) \right],
$$
we can write the solution to Eq. \eqref{equation_delta2} in the following compact form
\begin{equation}
\label{solution1}
\phi_{1}(x,t)=\exp\left( t\mathcal{L}\right)(\varphi(x)-\phi^e_1(x)) +\phi^e_1(x).
\end{equation}
Here, $\exp\left( t\mathcal{L}\right)$ is the time-dependent action of the diffusion operator on 
the initial density $\varphi(x)-\phi^e_1(x)$ while preserving the absorbing boundary conditions. This operator 
can be diagonalized in the basis of Gegenbauer polynomials on $L^2([0,1])$; see 
\cite{kimuraI}. The corresponding eigenvalues of $\exp\left( t\mathcal{L}\right)$ are
$\exp(-(i+1)(i+2)t/4N_e)$ with $i\in [0,\, \infty)$.

We can solve Eq. \eqref{equation_exp} in a similar way, by using the decomposition
\begin{equation}
\phi_{2}(x,t)=\phi_2^{h}(x,t)+\phi^e_2(x).
\end{equation}
In this case, $\phi^e_2(x)$ is the equilibrium density associated with the exponential mutation density,
as defined in Eq. \eqref{eq_exp}. The term $\phi_2^{h}(x,t)$ evolves under pure random drift, with no 
mutation influx, and initial condition $\phi^{h}_2(x,t=0)=\varphi(x)-\phi^e_2(x)$:
\begin{equation}
\label{solution2}
\phi_{2}(x,t)=\exp\left( t\mathcal{L}\right)(\varphi(x)-\phi^e_2(x)) +\phi^e_2(x).
\end{equation}
By subtracting Eq. \eqref{solution2} from Eq. \eqref{solution1}, we can describe the time evolution of the 
deviation as
\begin{equation}
\label{deviation_evolution}
\phi_{1}(x,t)-\phi_{2}(x,t)=-\exp\left( t\mathcal{L}\right)(\phi^e_1(x)-\phi^e_2(x)) +\phi^e_1(x)-\phi^e_2(x),
\end{equation}
which is independent of the initial condition $\varphi(x)$. 

One can show that $\phi^e_1(x)-\phi^e_2(x)$ is non negative on $[0,1]$, if $\kappa\leq 2N_e$. This can be seen more 
clearly by computing $\phi^e_1(x)-\phi^e_2(x)$ in the large $\kappa$ limit
\begin{eqnarray}
\phi^e_1(x)-\phi^e_2(x)=\frac{4N_e u}{1-x}(2N_e-\kappa), \quad\quad x\in [0,1/(2N_e)), \label{conditionI} \\
\phi^e_1(x)-\phi^e_2(x)=\frac{4N_e u e^{-\kappa x}}{x(1-x)}+O(e^{-\kappa}/(1-x)), \quad\quad x\in (1/(2N_e),1]. \label{conditionII}
\end{eqnarray}
The terms of order $e^{-\kappa}$ in Eq. \eqref{conditionII} exactly cancel the divergence at $x=1$.
Therefore, the action of the diffusion operator on $\phi^e_1(x)-\phi^e_2(x)$, will preserve the non-negativity
of the density
\begin{equation}
\exp\left( t\mathcal{L}\right)(\phi^e_1(x)-\phi^e_2(x)) \geq 0, \quad\quad \forall x\in [0,1],\quad \forall t>0.
\end{equation}
Because of this inequality, the absolute value $\vert \exp\left( t\mathcal{L}\right)(\phi^e_1(x)-\phi^e_2(x))\vert$, is the same
as $\exp\left( t\mathcal{L}\right)(\phi^e_1(x)-\phi^e_2(x))$, and we can evaluate exactly the integral
\begin{equation}
\int_{0}^1 \vert \exp\left( t\mathcal{L}\right)(\phi^e_1(x)-\phi^e_2(x))\vert x(1-x)dx=
\int_{0}^1 \exp\left( t\mathcal{L}\right)(\phi^e_1(x)-\phi^e_2(x)) x(1-x)dx, \label{l2projection}
\end{equation}
by expanding $\exp\left( t\mathcal{L}\right)(\phi^e_1(x)-\phi^e_2(x))$ in the eigenbasis of 
$\exp\left( t\mathcal{L}\right)$. This basis is orthogonal under the $L^2$-product
defined by the weight $x(1-x)$, and the constant function on $[0,1]$ corresponds to
the eigenfunction with the smallest eigenvalue. 
In this way we can interpret the right-hand side of Eq. \eqref{l2projection} 
as a projection on this eigenfunction, and evaluate the integral exactly. 

The eigenbasis of $\exp\left( t\mathcal{L}\right)$ is defined by the Gegenbauer polynomials.
As an example, the first three Gegenbauer polynomials on $[0,1]$, orthonormal under the 
$L^2$-product with weight $x(1-x)$, are 
\begin{eqnarray}
T_0(x) = \sqrt{6},\\
T_1(x) = \sqrt{30}(1 - 2x),\\
T_2(x) = \sqrt{84}(1 - 5x + 5x^2).
\end{eqnarray}
The corresponding eigenvalues in $\exp\left( t\mathcal{L}\right)$, are
eigenvalues $\exp(-t/(2N_e))$, $\exp(-3t/(2N_e))$, and $\exp(-3t/N_e)$. Thus, 
Eq. \eqref{l2projection} is the same as
\begin{equation}
\int_{0}^1 \exp\left( t\mathcal{L}\right)(\phi^e_1(x)-\phi^e_2(x))x(1-x)dx=
\int_{0}^1 \exp\left( t\mathcal{L}\right)(\phi^e_1(x)-\phi^e_2(x))\frac{T_0(x)}{\sqrt{6}} x(1-x)dx,
\end{equation}
and
\begin{equation}
\exp(-t/(2N_e)) \int_{0}^1 (\phi^e_1(x)-\phi^e_2(x))x(1-x)dx =\frac{4N_e u}{\kappa}\exp(-t/(2N_e)) +O(e^{-\kappa}).
\end{equation}
As $0\leq \exp\left( t\mathcal{L}\right)(\phi^e_1(x)-\phi^e_2(x)) \leq \phi^e_1(x)-\phi^e_2(x)$ for all $x\in [0,1]$ and
for $t\gg N_e$, we lastly compute Eq. \eqref{deviation_a}, as
\begin{eqnarray}
\int_{0}^1 \vert  \phi_1(x,t) -\phi_2(x,t)\vert x(1-x)dx = 
\int_{0}^1 ( \phi^e_1(x)-\phi^e_2(x)) x(1-x)dx \nonumber \\- 
\int_{0}^1 \exp\left( t\mathcal{L}\right) ( \phi^e_1(x)-\phi^e_2(x)) x(1-x)dx,\label{deviation_aa}
\end{eqnarray}
which is
\begin{equation}
\label{deviation_aaa}
\int_{0}^1 \vert  \phi_1(x,t) -\phi_2(x,t)\vert x(1-x)dx = 
\frac{4N_e u}{\kappa}\left( 1-\exp(-t/(2N_e))\right) +O(e^{-\kappa}),
\end{equation}
as we wanted to show.

\newpage

\bibliographystyle{plain}

\end{document}